\documentclass[a4paper,fleqn,usenatbib]{mnras}

\usepackage{mathptmx}

\usepackage[T1]{fontenc}
\usepackage{ae,aecompl}

\newcommand{\bc}{}
\usepackage{graphicx}	
\usepackage{amsmath}	
\usepackage{amssymb}	

\title[Testing exoplanet evaporation]{Testing exoplanet evaporation with multi-transiting systems}

\author[Owen, J. E. \& Campos Estrada, B.]{
James E. Owen\thanks{E-mail: james.owen@imperial.ac.uk}
 and Beatriz Campos Estrada
\\
Astrophysics Group, Department of Physics, Imperial College London, Prince Consort Rd, London, SW7 2AZ, UK
}

\pubyear{2015}

\begin{document}
\label{firstpage}
\pagerange{\pageref{firstpage}--\pageref{lastpage}}
\maketitle

\begin{abstract}
The photoevaporation model is one of the leading explanations
for the evolution of small, close-in planets and the origin of the radius-valley. However, without planet mass measurements, it is challenging to test the photoevaporation scenario. Even if masses are available for individual planets, the host star's unknown EUV/X-ray history makes it difficult to assess the role of photoevaporation.  We show that systems with multiple transiting planets are the best in which to rigorously test the photoevaporation model. By scaling one planet to another in a multi-transiting system, the host star's uncertain EUV/X-ray history can be negated. By focusing on systems that contain planets that straddle the radius-valley, one can estimate the minimum-masses of planets above the radius-valley (and thus are assumed to have retained a voluminous hydrogen/helium envelope). This minimum-mass is estimated by assuming that the planet below the radius-valley entirely lost its initial hydrogen/helium envelope, then calculating how massive any planet above the valley needs to be to retain its envelope. We apply this method to 104 planets {\bc above the radius gap} in 73 systems for which precise enough radii measurements are available. We find excellent agreement with the photoevaporation model. Only two planets ({\it Kepler} - 100c \& 142c)  appear to be inconsistent, suggesting they had a different formation history or followed a different evolutionary pathway to the bulk of the population. Our method can be used to identify {\it TESS} systems that warrant radial-velocity follow-up to further test the photoevaporation model. 
\\
\\
The software to estimate minimum planet masses is publicly available at:  \hyperlink{https://github.com/jo276/EvapMass}{https://github.com/jo276/EvapMass}
\end{abstract}



\section{Introduction}
Recent exoplanet discoveries have fundamentally changed our understanding of what typical planets and planetary systems are. These recent discoveries have been driven by both transit searches, such as the {\it Kepler} mission and precision radial-velocity (RV) measurements \citep[e.g.][]{Borucki2011,Mayor2011,Thompson2018}. We now know planets with radii in the range $1-4$~R$_\oplus$ and orbital periods $<100$~days are incredibly common \citep[e.g.][]{Howard2010,Fressin2013,Silburt2015,Mulders2018,Zink2019}, yet the formation of these close-in super-Earths/mini-Neptunes is still poorly understood \citep[e.g.][]{Jankovic2019}. The success of transiting searches for exoplanets means that there are thousands of confirmed exoplanets; yet, the vast majority of these do not possess a mass measurement. 

However, the combination of RV follow-up \citep[e.g.][]{Weiss2014,Marcy2014} and transit-timing variations \citep[TTVs, e.g.][]{Wu2013,Hadden2014,Xie2014,JontofHutter2016,Hadden2017} for a small fraction of transiting exoplanets has allowed some constraints on density and hence composition of exoplanets. Planets with radii $\lesssim 2$~R$_\oplus$  typically have densities which imply they have a composition similar to Earth \citep[e.g.][]{Dressing2015,Dorn2019}. Whereas larger planets have lower densities, implying that they must contain a significant fraction of volatiles \citep[e.g.][]{Rogers2010,Weiss2014,Rogers2015}, in many cases the densities are so low that their densities can only be explained if the planets contain voluminous H/He atmospheres \citep[e.g.][]{JontofHutter2016}. With ongoing transit searches expected to return thousands more exoplanets \citep[e.g.][]{Gunther2017,Barclay2018} in the next few years, the fact that the majority of known exoplanets will only have measured radii is likely to remain for the foreseeable future. Therefore, it is crucial that we develop techniques and methods to constrain planetary masses and compositions without direct measurements of planet mass. 

Planets with H/He atmospheres on short period orbits are liable to mass-loss due to photoevaporation \citep[e.g.][]{Lammer2003,MurrayClay2009,Owen2012}. High-energy irradiation (X-ray and UV) heats up the upper-layers of the H/He atmosphere driving a powerful hydrodynamic outflow that causes the atmosphere to lose mass over time \citep[e.g.][]{Baraffe2005,Lopez2012,Owen2019}, and in some cases completely removing it. {\bc Such mass-loss has been observed to be occurring through transmission spectroscopy. Originally, this was done using the Lyman-$\alpha$ line \citep[e.g.][]{VidalMadjar2003,Lecavelier2010,Ehrenreich2015} and more recently in the 10830\AA~ HeI line \citep[e.g.][]{Spake2018,Allart2018}.}  This idea has led to the hypothesis studied by \citet{Owen2013} and \citet{Lopez2013} that the majority of close-in, low-mass planets are born with a composition of a solid core surrounded by a low-mass, but voluminous, H/He atmosphere which then experiences photoevaporation and mass-loss. \citet{Owen2013} demonstrated that this evolutionary pathway resulted in two distinct planetary structures after billions of years of evolution: firstly, lower mass, more highly irradiated planets typically completely lose their H/He atmosphere and finish as a ``stripped'' core; secondly, higher-mass, less irradiated planets typically end up with a H/He atmosphere that consists of $\sim1$\% of the planet's mass, but the radius of the planet is double that of the core. These distinct evolutionary pathways result in a gap in the radius and radius-period distribution of close-in exoplanets and in a distinct prediction of this hypothesis for the origin and evolution of close-in exoplanets, having been confirmed in subsequent theoretical works \citep[e.g.][]{Lopez2013,Jin2014,Chen2016}.

A gap has now been observed in the radius distribution, where there is a distinct lack of planets with radii $\sim 1.8$~R$_\oplus$ \citep{Fulton2017,Fulton2018} in agreement with the photoevaporation model. Furthermore, using a set of planets with precise parameters determined through asteroseismology, \citet{VanEylen2018} demonstrated that this gap is clean, and declines with period in excellent agreement with the photoevaporation model. As argued by \citet{Owen2013} and demonstrated in \citet{Owen2017}, with an observed radius-gap one can use the photoevaporation model to infer the mass-distribution and hence core-composition of close-in exoplanets. Comparisons to the exoplanet data using the photoevaporation-driven evolution model find that the core composition of these planets is iron-rich and ``Earth-like'' \citep{Owen2017,Jin2018,Wu2019}, providing challenging constraints on the formation of close-in super-Earths and mini-Neptunes. 

However, conclusions about the formation pathways and compositions of close-in exoplanets are prefaced on the photoevaporation model being correct and the mass-loss rates being accurate. There is a degeneracy between the derived core-composition and photoevaporative mass-loss rates, with lower mass-loss rates favouring lower core densities \citep{Wu2019,OA19}. Furthermore, alternative hypothesis have been suggested for the origin of the observed gap in the radius distribution, including core-powered mass-loss \citep{Ginzburg2018,Gupta2019b,Gupta2019}, or two distinct formation pathways for the two sub-groupings. In the latter scenario, the two sub-groupings are water-worlds and rocky, terrestrial planets \citep{Zeng2019}. {\bc Recently, it has also been suggested planetesimal impacts may create a similar radius-gap \citep{Wyatt2019}.}

The uncertainty over the formation and evolutionary history of close-in exoplanets stems from the fact that the models are under constrained, as the exoplanet mass function remains unknown, but is rather inferred from the chosen evolutionary model. These different models infer different mass functions. In addition, for the photoevaporation model, the majority of the evolution occurs within the first $\sim 100$~Myr of the star's life when its UV \& X-ray output is considerably higher, and the star's current high-energy output is not representative of its earlier history, with at least an order of magnitude spread possible \citep[e.g.][]{Tu2015}. This means that comparisons of individual planets to the photoevaporation models are weakly constraining as one does not know an individual star's high-energy output over it's lifetime. 

However, in multi-planet systems while the star's high-energy output is still uncertain we know that all planets in the system experienced the same history. This means multi-planet systems provide an excellent test bed for the photoevaporation model, as has already been demonstrated for the {\it Kepler}-36 system \citep{Lopez2013,Owen2016}. In this work, we argue that multi-planet systems which contain planets both above and below the radius-gap (i.e. they ``straddle'' the gap) are particularly powerful for testing the photoevaporation model. If one adopts the photoevaporation model then the current few billion-year old architecture of the multi-planet systems allows constraints to be placed on the minimum masses of planets above the radius-gap in order to be consistent with the photoevaporation model. Specifically, we ask the question: what is the minimum mass a planet must have in order to retain its H/He atmosphere, given another planet in the system entirely lost one?

\section{Concept and assumptions}

Before we describe our method in detail it is useful to outline the concept of how we can use photoevaporation to constrain planetary masses and the assumptions on which it is based. In order for this method to be applicable we require a multi-transiting exoplanet system that contains at least one super-Earth that is below the radius gap and at least one mini-Neptune with a radius above the radius gap (we use the nomenclature of a super-Earth planet being a planet below the radius-gap and a mini-Neptune being above the radius-gap throughout this work). Although, we caution that the method is not applicable to planets where the mass in the H/He atmosphere is comparable to or larger than the mass in the core, wherein self-gravity of the planet's atmosphere becomes important. Therefore, we crudely apply a cut in planetary radii of $<6~$R$_\oplus$. 

In order to proceed we then assume that the super-Earth planet was born with a H/He atmosphere which is then lost. We take the super-Earth planet to have {\it just} been able to lose any initial H/He atmosphere, hence maximising its mass-loss timescale. Then we solve for the core-mass that equates this {\it maximum} mass-loss timescale to the mass-loss timescale for the mini-Neptune to {\it just} retain its current H/He atmosphere. This procedure minimises the mass-loss timescale for the mini-Neptune, hence placing a minimum constraint on the core-mass of the mini-Neptune to be consistent with the photoevaporation model. This is because higher core masses will have longer mass-loss timescales and will also be consistent with the photoevaporation models. 
\subsection{Assumptions}\label{sec:assump}
The assumptions are based on the general conclusions obtained by \citet{Owen2017}, \citet{Jin2018} and \citet{Wu2019} when applying the photoevaporation model to the exoplanet radius data alone. The assumptions are as follows:
\begin{enumerate}
    \item We assume that the core-composition of all planets in the multi-transiting systems are identical.
    \item We assume that all planets have remained on the currently observed orbits since disc dispersal.
    \item  We assume that all planets accreted a H/He atmosphere from the protoplanetary disc that initially contained an atmospheric mass $\gtrsim 1$\%.
    \item  Finally, in addition to the first assumption, we further assume that the composition of the cores is ``Earth-like'' containing 1/3 iron and 2/3 silicate rocks by mass. 
\end{enumerate}

It is important to emphasise that none of these assumptions can be convincingly argued from a first-principled approach to planet formation. Rather these assumptions have been {\emph{inferred}}, by comparing the photoevaporation model to the exoplanet data.
Now any planet that has a measured mass below that required to be consistent with the photoevaporation model is likely to result from the breaking of one of the above assumptions. Such a comparison would allow the identification of planetary systems that could have undergone giant impacts after the disc disperses \citep[e.g.][]{Inamdar2016}, or have ice-rich cores, or those with variable core compositions \citep[e.g.][]{Raymond2018}. Finally, if the measured masses are typically below those required to be consistent with the photoevaporation model, then the model can be ruled out as the origin of the observed radius gap.  

\subsection{Basic Expectations}\label{sec:basic}
Before we proceed with numerical solutions, we can get a basic expectation of how the derived minimum mass depends on the parameters of the multi-planet system using the analytic scalings of \citet{Owen2017}. We adopt the energy-limited mass-loss formula, $\dot{m} = \eta\pi R_p^3 L_{\rm HE} / 4 \pi a^2 G M_p$, with $\eta$ being the mass-loss efficiency, $R_p$, $M_p$ and $a$ the planet's mass, radius and separation,  $L_{\rm HE}$ the high energy luminosity of the star during its active period {\bc and $G$ the gravitational constant}. In this sub-section, we take the maximum mass-loss timescale for the planet's to occur for an envelope mass fraction $X_2$ that doubles the core's radius, i.e. $R_p=2 R_c$, with $R_c$ the core's radius.  Therefore, writing the mass-loss timescale $t_{\dot{m}}\equiv M_{\rm atm}/\dot{m}$, we find the maximum mass-loss timescale is approximately:
\begin{equation}
    t^{\rm max}_{\dot{m}}\approx \frac{a^2GM_p^2X_2}{\eta 2R_c^3L_{HE}}
\end{equation}
We now take the above expression to represent the maximum mass-loss timescale for the super-Earth. We can now say the mass-loss timescale for the mini-Neptune must be larger than this in order to remain gaseous. Crudely, also evaluating this at the point of maximum mass-loss timescale for the mini-Neptune (i.e. when the atmosphere mass fraction is $X_2$), we can then find a constraint on the minimum mass of the mini-Neptune. This gives:
\begin{eqnarray}
t^{\rm gas}_{\dot{m}}&\geq& t^{\rm rock}_{\dot{m}}\nonumber\\
\frac{a^2_gM_{g}^2}{\eta_g R_{c,g}^3}&\geq&\frac{a^2_RM_{R}^2}{\eta_R R_{c,R}^3}\nonumber \\
M_g&\geq& M_R\,\left(\frac{a_R}{a_g}\right)\left(\frac{\eta_g}{\eta_R}\right)^{1/2}\left(\frac{R_{c,g}}{R_{c,R}}\right)^{3/2}\label{eqn:simple1}
\end{eqnarray}
where the sub-script $R$ refers to the ``rocky'' super-Earth and $g$ refers to the ``gaseous'' mini-Neptune. Since both mass-loss timescales depend inversely on $L_{HE}$ this minimum mass is independent of the unknown high-energy flux history of the star. Now adopting the simple mass-radius relation for the solid cores ($M_c\propto R_c^4$, e.g., \citealt{Valencia2007}), we can further simplify Equation~\ref{eqn:simple1} to become:
\begin{eqnarray}
M_g&\geq& M_R\, \left(\frac{a_R}{a_g}\right)^{8/5}\left(\frac{\eta_g}{\eta_R}\right)^{4/5}\nonumber\\
M_g&\geq& 5.1\,M_\oplus \left(\frac{R_R}{1.5\,{\rm R}_\oplus}\right)^{4}\left(\frac{a_R}{a_g}\right)^{8/5}\left(\frac{\eta_g}{\eta_R}\right)^{4/5}
\end{eqnarray}
were the last inequality has been evaluated for Earth-like composition cores, in terms of the observable radius of the planet below the radius gap. Now finally, if one adopts the scaling of the mass-loss efficiency with escape velocity as $\eta \propto v_{\rm esc}^{-2}$ from \citet{Owen2017}, we find a simple relation between the minimum mass of the gaseous planet and the ratio of the orbital separations as: 
\begin{equation}
M_g\geq 5.1\,M_\oplus \left(\frac{R_R}{1.5\,{\rm R}_\oplus}\right)^{4}\left(\frac{a_R}{a_g}\right) \label{eqn:simple2}
\end{equation}

Unsurprisingly, one can see the most constraining systems for the photoevaporation model are those that contain a large super-Earth, or those that contain a super-Earth that is exterior to the mini-Neptune. But, those multi-planet systems that contain a small super-Earth interior to a mini-Neptune are unlikely to be strongly constraining. 

As a simple demonstration we can apply the above model to the {\it Kepler}-36 system \citep{Carter2012}, which contains two planets which straddle the gap. Planet b has a radius of 1.49~R$_\oplus$ and a separation of 0.l153~AU while planet c has a radius of 3.68~R$_\oplus$ and a separation of 0.1283~AU \citep{Carter2012}. Therefore, Equation~\ref{eqn:simple2} implies in order to be consistent with the photoevaporation model planet c should have a mass $\gtrsim 4.6$~M$_\oplus$. This minimum mass is consistent with its measured mass of $\sim 8.1$~M$_\oplus$ \citep{Carter2012}, implying the current observed architecture of the {\it Kepler}-36 system is consistent with the photoevaporation scenario. This result is not surprising given detailed studies of the {\it Kepler}-36 system have already shown it to be in agreement with the photoevaporation scenario \citep[e.g.][]{Lopez2013,Owen2016}.

\section{Overview of the Method}\label{sec:method}
While the discussion and derivation in Section~\ref{sec:basic} was a useful prelude to demonstrate the idea, the resulting Equation~\ref{eqn:simple2} neglected several key aspects that need to be included before we can robustly compare the photoevaporation model to real systems. 

Specifically, we assumed that the envelope-mass fraction at which the mass-loss timescale was maximised is independent of planetary properties. Furthermore, we took the gaseous planet to also have an envelope mass-fraction which maximises its mass-loss timescale. Such an assumption while useful, and true on average, is not true for specific planets which may have envelope mass fractions larger than that required to maximise the mass-loss timescale. An envelope-mass fraction larger than this will result in a shorter envelope-mass fraction and require a larger planet mass to compensate. We also need to account for the fact that planet's H/He atmospheres contract over time and were larger when mass-loss was important compared to when we observe them today. Finally, we also need to account for observational errors in the planetary properties. 

However, the basic procedure remains the same. We are solving for the mass-loss timescale of the gaseous planet such that it is greater than or equal to the maximum mass-loss timescale the rocky planet would have had for any atmosphere mass-fraction (provided it's $<$1).  

Therefore, the first goal is to find the maximum mass-loss timescale for the now supposed stripped core and the envelope mass-fraction that maximises it. Namely, we are maximising the ratio:
\begin{equation}
    \frac{X(M_p,a)}{\eta(R_p,M_p,a)R_p(X,M_p,a)^3}\label{eqn:max_rocky}
\end{equation}
with $X$ now the general envelope mass fraction $X\equiv M_{\rm env}/{M_{\rm core}}$. We assume that $M_p=M_c$ and $M_c$ is obtained from a mass-radius relation for the solid core, for which we use the \citet{Fortney2007} relations. {\bc In this work we have chosen to adopt the simple mass-loss efficiency of \citet{Owen2017}. However, the model can equally be applied to any mass-loss model, for example detailed numerical radiation-hydrodynamic simulations \citep[e.g.][]{Owen2012,Kubyshkina2018}. In-fact finding planetary systems that are inconsistent with our test may point, not to errors in the underlying photoevaporation scenario, but rather errors in the mass-loss efficiency. }

\subsection{Envelope Structure Model}
\begin{figure*}
\centering
\includegraphics[width=\textwidth]{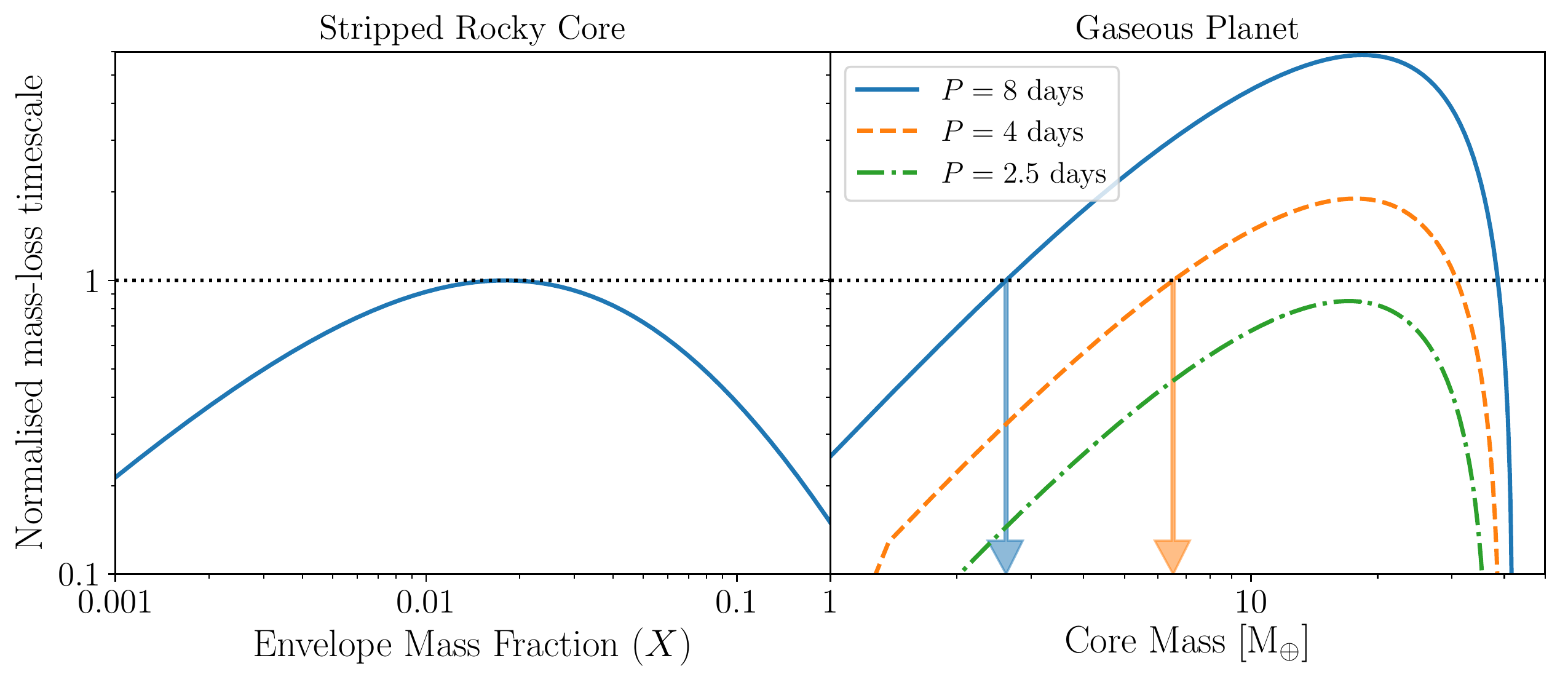}
\caption{The left panel shows the mass-loss timescale as a function of envelope mass fraction for a 1.5~R$_\oplus$ planet with a period of 5~days, which we assume has been striped by photoevaporation. The mass-loss timescale is normalised to its maximum. The right panel shows the mass-loss timescale as a function of core-mass for a 2.5~R$_\oplus$ planet, which we assume has retained a H/He envelope at periods of 2.5, 4 and 8~days. By equating the maximum mass-loss timescale for the assumed striped rocky core to the assumed gaseous planet we can find the minimum core mass of the gaseous planet to be consistent with the photoevaporation scenario (shown by the arrows). We find a minimum mass of 2.62~M$_\oplus$ if the gaseous planet had an orbital period of 8~days and 6.53~M$_\oplus$ if it had an orbital period of 4~days and no solution consistent with the photoevaporation model if it had an orbital period of 2.5~days.    }\label{fig:compare}
\end{figure*}
In order to convert the envelope mass fraction into a planetary radius or vice-versa we need an envelope structure model. We use the simple semi-analytic envelope structure model adopted by \citet{Owen2017} where the envelope consists of an adiabatic interior {\bc (with adiabatic index $\gamma$)} from the surface of the core to a radiative-convective boundary at radius $R_{\rm rcb}$ {\bc and density $\rho_{\rm rcb}$}, followed by an isothermal radiative layer at the equilibrium temperature {\bc ($T_{\rm eq}$)} reaching the photosphere at the planet's radius \footnote{\bc It is of course an assumption that the interior is convective; however, at this stage such a choice is the natural starting point.}. We choose this semi-analytic approach rather than solving for the exact structure (e.g. by using {\sc mesa}, \citealt{Paxton2011,Paxton2013}) as it is computationally inexpensive, this allows us to include errors in the planetary parameters by Monte-Carlo sampling (Section~\ref{sec:errors}) and quickly analyse a large number of planetary systems.   

The density profile in the convective interior is adiabatic and can be approximated by\footnote{This assumes that the mass contained in the envelope near the radiative convective boundary is negligible, this approximation breaks down when the radiative convective boundary is thin.}:
\begin{equation}
    \rho\approx\rho_{\rm rcb}\left[\nabla_{\rm ab}\left(\frac{GM_c}{c_s^2R_{\rm rcb}}\right)\left(\frac{R_{\rm rcb}}{r}-1\right)\right]^{1/(\gamma-1)}
\end{equation}
{\bc where $\nabla_{\rm ab}$ is the adiabatic gradient, $c_s$ is the isothermal sound-speed, and $r$ is the radius from the centre of the core.} This allows the mass in the convective interior to be written as:
\begin{equation}
    M_{\rm env}\approx 4\pi R_{\rm rcb}^3\rho_{\rm rcb}\left(\nabla_{\rm ab}\frac{GM_c}{c_s^2R_{\rm rcb}}\right)^{1/(\gamma-1)}I_2(R_c/R_{\rm rcb},\gamma)
\end{equation}
where $I_n$ is a dimensionless integral of the form:
\begin{equation}
    I_n(R_c/R_{\rm rcb},\gamma)=\int_{Rc/R_{\rm rcb}}^1 x^n\left(x^{-1}-1\right)^{1/(\gamma-1)}{\rm d}x
\end{equation}

The density profile in the isothermal radiative layer can be approximated by:
\begin{equation}
    \rho=\rho_{\rm rcb}\exp\left(-\frac{R-R_{\rm rcb}}{H}\right)\label{eqn:iso_atm}
\end{equation}
with $H=c_s^2R_{\rm rcb}^2/G M_p$ the isothermal scale height. Note, Equation~\ref{eqn:iso_atm} assumes the isothermal radiative atmosphere is thin, and the gravitational acceleration is constant. Additionally, in most cases we assume that the envelope mass is entirely contained within the convective interior and the isothermal radiative layer does not contribute to the envelope mass.  Finally, in order to evaluate the density at the radiative convective boundary we equate the temperature gradient across the radiative-convective boundary. \citet{Owen2017} demonstrate that for an opacity law of the form $\kappa=\kappa_0 P^{\alpha}T^{\beta}$ this gives the density at the radiative convective boundary of:
\begin{equation}
    \rho_{\rm rcb}\approx \left(\frac{mu}{k_b}\right)\left[\left(\frac{I_2}{I_1}\right)\frac{64\pi\sigma T_{\rm eq}^{3-\alpha-\beta}R_{\rm rcb}\tau_{\rm KH}}{3\kappa_0M_pX}\right]^{1/(1+\alpha)}
\end{equation}
with $\tau_{\rm KH}$ the Kelvin-Helmholtz timescale of the convective interior, which we equate to the age of the planet, {\bc and $\sigma$ the Stefan-Boltzmann constant}. Following, \citet{Owen2017} and \citet{Wu2019} we select $\gamma=5/3$, $\alpha=0.68$, $\beta=0.45$ and finally, $\kappa_0=4.79\times10^{-8}$ when pressure and temperature are expressed in cgs units. 

\subsection{Calculation of the minimum planetary mass}
The dependence of the planetary structure on age (i.e. younger planets have lower mass envelopes compared to the older planets with the same radius), means that we cannot quite ignore the age at which mass-loss is important and the current age of the planets. Although, as we show in Section~\ref{sec:results} this trend is weak. Therefore, we select two timescales, $t_{\rm young}=100$~Myr, which is when we equate the two mass-loss timescales and $t_{\rm old}=\,$system age, which is when we compare to the observed radius of the supposedly gaseous planet. Therefore, when we find the maximum mass-loss timescale for the rocky planet in maximising Equation~\ref{eqn:max_rocky} we set $\tau_{\rm KH}=t_{\rm young}$ and we also equate this to the mass-loss timescale for the gaseous planet's atmosphere; however, we also constrain the gaseous planet's atmosphere to have an envelope mass-fraction consistent with the observed planet's radius at $t_{\rm old}$. This means the planetary radius that was adopted for the gaseous planet when it's mass-loss timescale is equated to the maximum mass-loss timescale for the rocky planet is slightly larger than the observed radius. 

\subsubsection{Demonstration Systems}
Before we apply our method to real observed planetary systems it is useful to demonstrate the approach for representative systems. As the method of equating the mass-loss timescales typically results in either two solutions for the mass of the gaseous planet or no solution, both outcomes are easy to understand. 

We consider a two planets system consisting of a 1.5~R$_\oplus$ planet (planet b) with an orbital period of 5~days and a second planet with a radius of 2.5~R$_\oplus$ (planet c) whose period we vary. We take planet b as a stripped rocky core (the super-Earth) and planet c as a planet that retained a H/He atmosphere (the mini-Neptune). Therefore, we wish to maximise the mass-loss timescale of planet b, assuming it had an atmosphere, this is shown in the left-hand panel of Figure~\ref{fig:compare}, where we plot the mass-loss timescale as a function of envelope mass fraction for planet b. This mass-loss timescale clearly peaks at an envelope mass-fraction of a few percent as expected. Next, for planet c we compute its mass-loss timescale as a function of core mass, requiring its radius to be 2.5~R$_\oplus$. Planet c's mass-loss timescale is shown in the right panel of Figure~\ref{fig:compare}. This mass-loss timescale is short at low core masses because any atmosphere is weakly bound to the core. It is also short at high core masses because the envelope mass fraction is small (i.e. as the core's radius approaches the planet's radius the envelope mass fraction approaches zero and the mass-loss timescale also tends to zero). 

Thus, when equating the maximum mass-loss timescale for planet b to planet c (see the right-panel of Figure~\ref{fig:compare}), we can clearly see that there is either two solutions, or no solutions. The two solutions are either a lower core mass with a substantial ($X\gtrsim$ 1\%) envelope mass fraction or at high core mass with very little envelope ($X\ll 1$\%). Now clearly, if planet c has a period of 2.5~days, all core-masses have a mass-loss timescale less than the maximum of planet b then there is no solution that is consistent with planet b having been completely stripped of a substantial envelope and planet c retaining one. 

Since the inference from the photoevaporation model, when compared to the radius data is that planets are born with a substantial atmosphere and we are interested in a minimum mass to be consistent with the photoevaporation model we select the lower mass solution, as shown by the arrows in the right-panel of Figure~\ref{fig:compare}, with closer in gaseous planets requiring higher core masses, as expected from our discussion in Section~\ref{sec:basic}. 

In our numerical method to ensure we find the correct solution we maximise the mass-loss timescale for the gaseous planet and then use a bounded root finder to search for a solution between the maximised core mass and a lower bound. In our analysis we set this lower bound to 0.1~M$_\oplus$. Therefore, any mini-Neptune whose mass-loss timescale is larger than the super-Earth for 0.1~M$_\oplus$ are just assigned a minimum mass of $<$~0.1~M$_\oplus$.  

\begin{figure}
    \centering
    \includegraphics[width=\columnwidth]{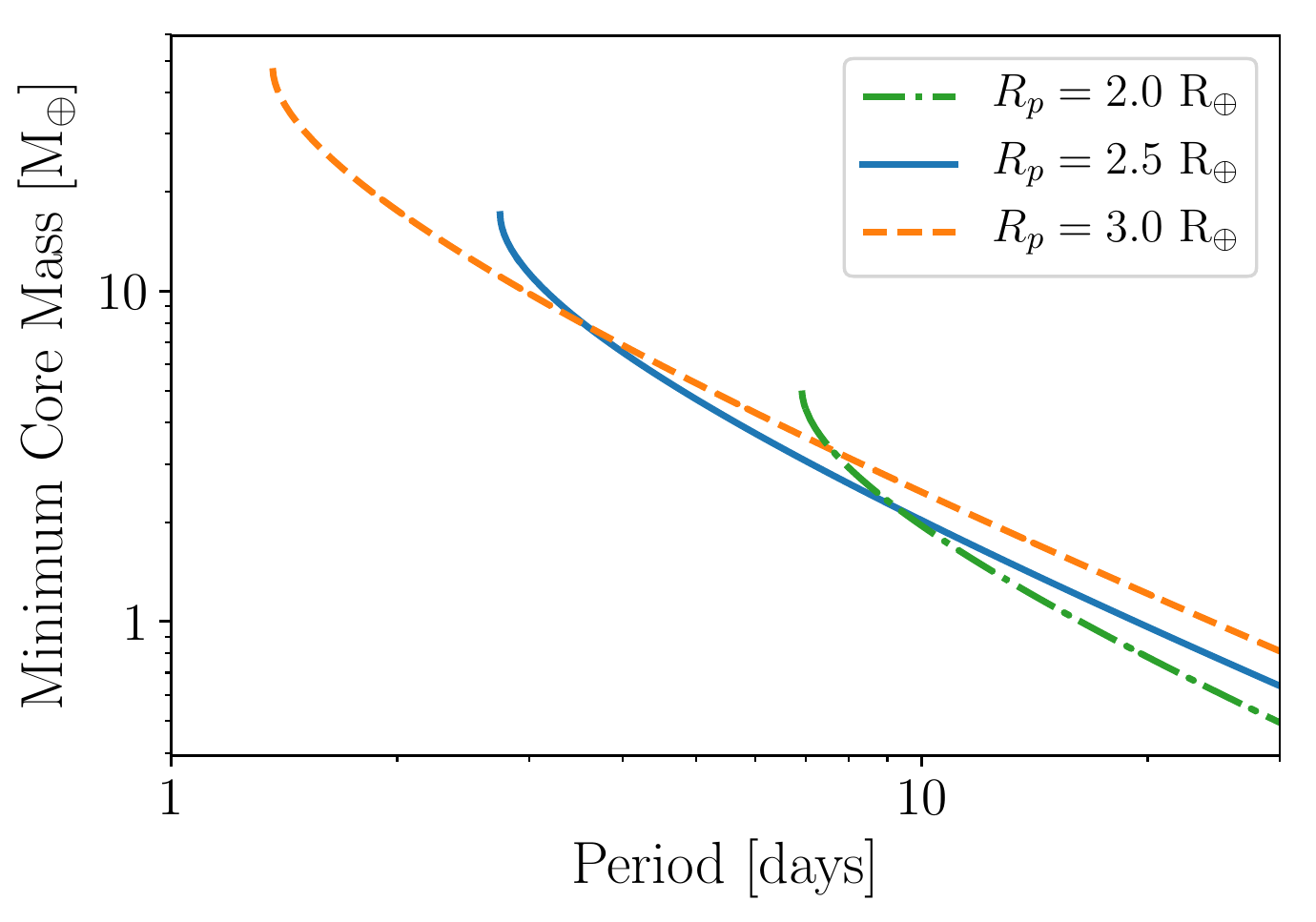}
    \caption{The minimum core mass to be consistent with photoevaporation for a mini-Neptune as a function of orbital period with a super-Earth in the same system that has a radius of 1.5~R$_\oplus$ and period of 5~days. This minimum core mass is shown for a mini-Neptune with radii of 2.0, 2.5 \& 3.0 R$_\oplus$. {\bc At long periods note the approximate $M_c\propto P^{-2/3}$ power-law, as expected from Equation~4.} }
    \label{fig:rad_vary}
\end{figure}

We can assess how our minimum mass estimate for planet c would vary if we change its radius. This is shown for planetary radii where we plot the minimum core mass as a function of period for planet c with a radius of 2.0, 2.5 \& 3.0 R$_\oplus$ in Figure~\ref{fig:rad_vary}. This shows, as expected, in general a larger planet, which one would expect to have a larger envelope mass-fraction requires a higher minimum core mass. However, larger planets can also be consistent closer to the star, where the larger planetary radii permit larger cores masses to fit inside that radius. For example, the $\gtrsim 20$~M$_\oplus$ minimum core masses that appear for the 3.0~R$_\oplus$ planet at short periods are not possible for the 2.0~$R_\oplus$ planet, since a $\sim 20~$M$_\oplus$ core is larger than 2.0~R$_\oplus$. 

\begin{figure}
    \centering
    \includegraphics[width=\columnwidth]{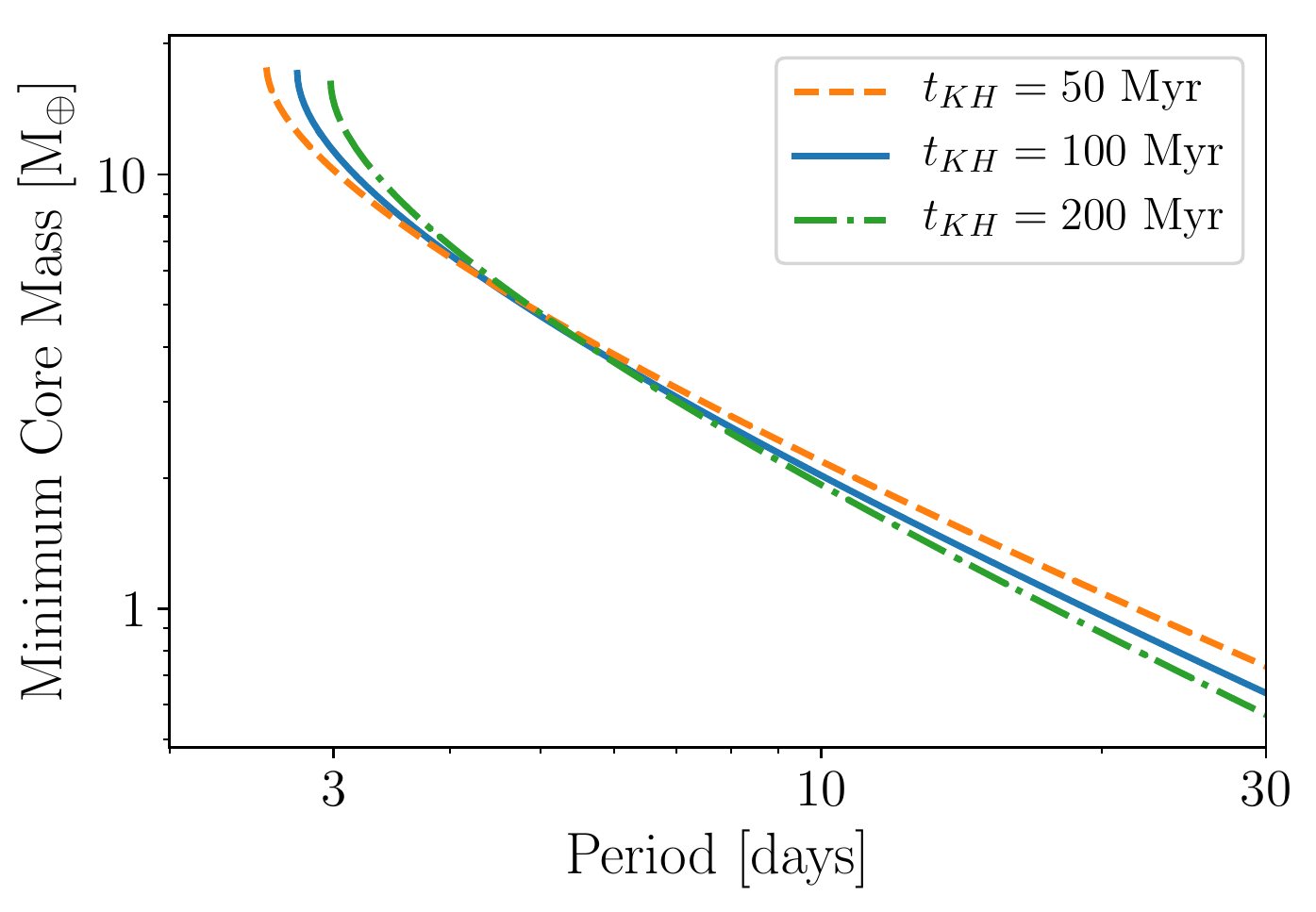}
    \caption{The minimum core mass to be consistent with photoevaporation for a mini-Neptune with a radius of 2.5~R$_\oplus$ as a function of orbital period with a super-Earth in the same system that has a radius of 1.5~R$_\oplus$ and period of 5~days. This minimum core mass is shown for three choices of the young planet's Kelvin-Helmholtz timescales ($t_{\rm young}$) of 50, 100, \& 200 Myr.  {\bc At long periods note the approximate $M_c\propto P^{-2/3}$ power-law, as expected from Equation~4.}}
    \label{fig:tkh_vary}
\end{figure}

Finally, in Figure~\ref{fig:tkh_vary} we assess how our choice of the age at which to evaluate the planetary structure affects our results. Here we plot the minimum core mass for planet c (with a radius of 2.5~R$_\oplus$) as a function of period for choices of $t_{KH}$ of 50,100 (our default choice) and 200~Myr. This shows, that the minimum estimated core masses do not strongly depend on this choice and such differences are likely to be small compared to those caused by errors in the planetary parameters (see Section~\ref{sec:errors}).   

\subsection{Inclusion of errors on planetary parameters}\label{sec:errors}
All of the key input parameters contain measurement errors and this must be folded into our analysis. Firstly, we must check that the multi-planet system robustly straddles the radius-gap. We do this by taking that the super-Earth's radius is below the gap to 2$\sigma$ and that the mini-Neptune's radius is above the gap to 2$\sigma$. 

In order to include errors on the minimum mass of the gaseous planet we assume Gaussian errors on all the planetary and stellar parameters that are independent\footnote{In reality one could use planetary parameters provided by MCMC chains from the transit fitting procedure, that include any co-variances between the planetary parameters.}. We then randomly draw from each of the planetary and stellar parameters and perform our minimum-mass estimate, we do this Monte Carlo sampling 3000 times and the minimum mass is then given as a 95\% upper-limit unless otherwise stated.

\section{Results}\label{sec:results}
Here we apply our method to observed exoplanet multi-transiting systems and then compare our estimated minimum masses to those systems that have constraints on their masses. 
\subsection{{\it Kepler}-36}
\begin{figure}
    \centering
    \includegraphics[width=\columnwidth]{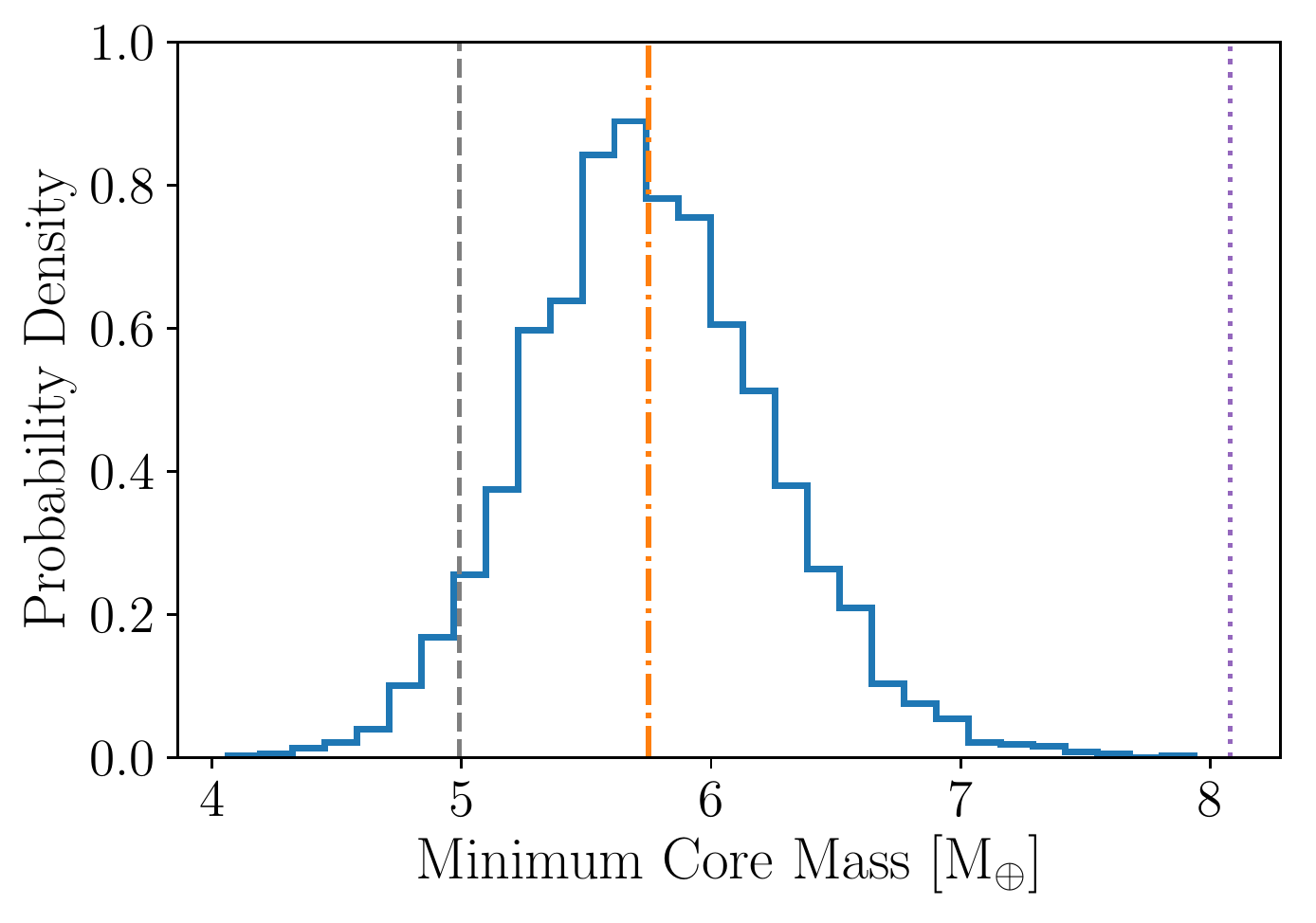}
    \caption{The distribution of minimum core-masses found for {\it Kepler}-36c. The dashed line gives the 95\% upper limit of 4.99~M$_\oplus$, the dot-dashed line gives the value if the evaluation just used the best fitting parameters for the {\it Kepler}-36 system and the dotted line shows best-fit measured mass for {\it Kepler}-36c from \citet{Carter2012}. }
    \label{fig:kepler36}
\end{figure}

We return again to the photoevaporative benchmark system {\it Kepler-36} and apply our full method to the observed system. We take the observed stellar and planetary parameters including errors from \citet{Carter2012}. Following the discussion in Section~\ref{sec:errors} we randomly sample the stellar and planetary parameters and calculate the minimum mass for each random draw. The resulting distribution for the minimum planetary mass of {\it Kepler-36} is shown in Figure~\ref{fig:kepler36}
which yields a 95\% upper-mass limit for the system to be consistent with photoevaporation of 4.99~M$_\oplus$, similar to the $\sim 4.6$~M$_\oplus$ upper limit we obtained from simple arguments in Section~\ref{sec:basic}, and as expected is consistent with its measured mass, indicating that the {\it Kepler}-36 system is consistent with the photoevaporative scenario and the assumptions listed in Section~\ref{sec:assump}. 

By artificially setting the errors on the stellar and planetary parameters to zero in turn, we find by far the most dominant source of spread in the minimum mass estimates is driven by the radius of the super-Earth {\it Kepler}-36b. With the radius of {\it Kepler}-36b fixed to its best-fit value the 95\% upper limit for the minimum mass of {\it Kepler}-36c is now $\sim$5.5~M$_\oplus$ only slightly smaller than the value obtained by just evaluating the best fit for all system parameters. This is not surprising, as the inferred mass of the rocky planet depends strongly (roughly $\propto R_p^4$) on its radius and the minimum mass for the gaseous planet scales linearly with the mass of the rocky one (Section~\ref{sec:basic}). Therefore, even small radius errors (which are 2.3\% for {\it Kepler}-36b) on the super-Earth planet dominate the precision on the minimum mass estimate. 

\subsection{Asteroseismic sample}\label{sec:astero}
\begin{figure}
    \centering
    \includegraphics[width=\columnwidth]{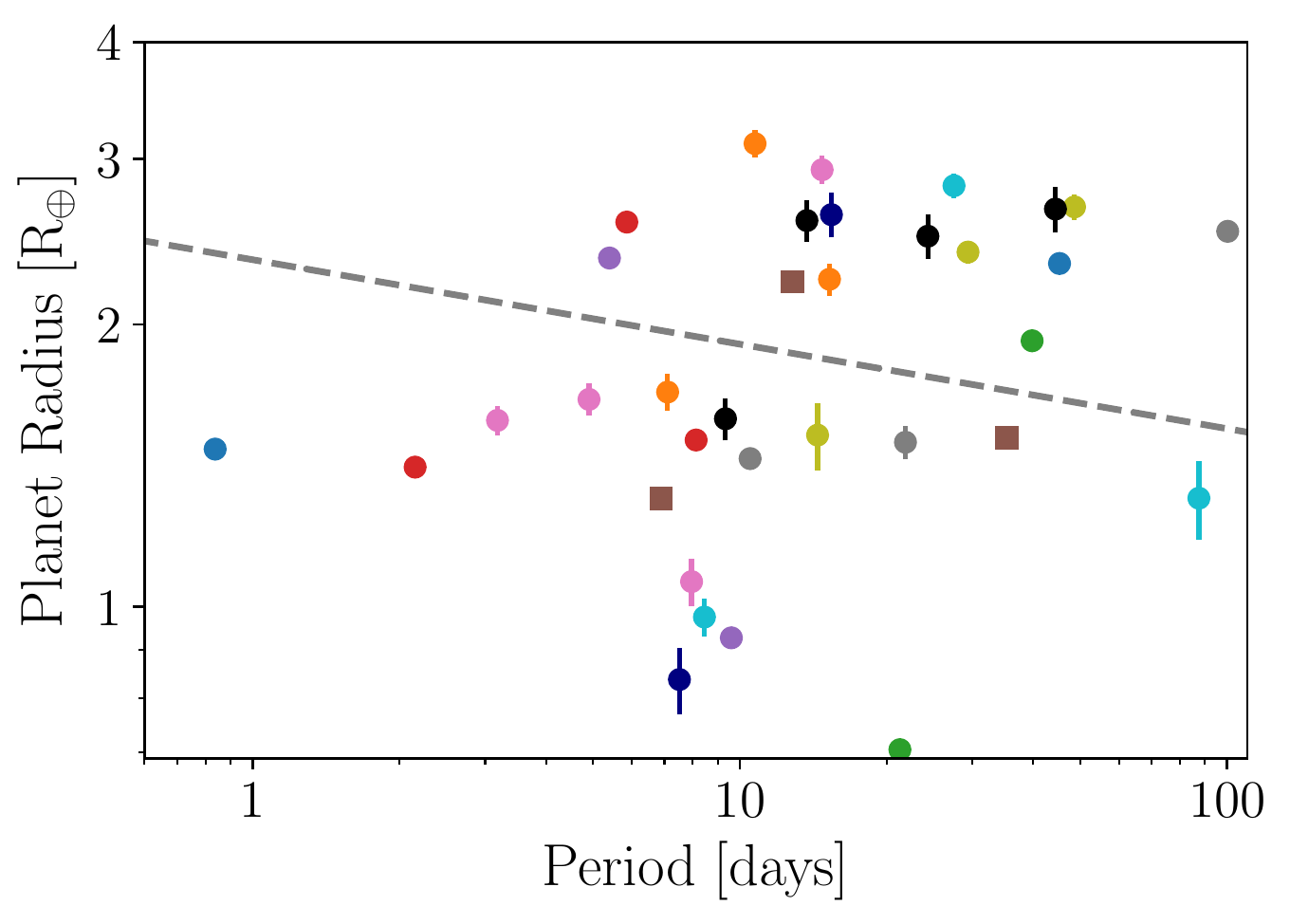}
    \caption{The radii and periods of planets from the asteroseismic sample in multi-transiting systems that straddle the radius gap. These 12 systems are analysed in Section~\ref{sec:astero} to derive the minimum masses for all the planets above the radius gap. Planets with the same colour indicate they are in the same system. Planets without visible error bars have errors smaller than the size of the symbols. The {\it Kepler}-100 system, which is found to be inconsistent with photoevaporation, is shown as square symbols rather than filled circles. The radius-gap derived by \citet{VanEylen2018} is shown as the dashed line.  }
    \label{fig:astero_multis}
\end{figure}
Here we work with the asteroseismic sample of \citet{VanEylen2018}, this set contains 24 multi-planet systems of which 13 straddle the radius-gap (for analysis in this sub-section we use the radius-gap as a function of period defined by \citealt{VanEylen2018} - {\bc we use their expression for the position determined using support vector machines - i.e. the line shown in their Figure~7}), 12 of which contain mini-Neptunes with radii $<6$~R$_\oplus$. These systems are shown in Figure~\ref{fig:astero_multis}. Eight of these systems have some published mass constraint (including upper-limits)\footnote{As listed in NASA's exoplanet archive, in the case of multiple masses listed we take the most recent values without prejudice.}. In total we compute mass estimates for 16 mini-Neptunes. The estimated minimum masses at the 95\% upper-limit level for these systems along with any mass-constraint, an estimated RV semi-amplitude for the minimum mass (assuming a circular orbit), $K_{\rm cir}$, and which super-Earth planet in the system these masses have been scaled from are listed in Table~\ref{tab:astero}. As expected from our earlier discussion those systems which are most constraining have a either a large and hence massive planet below the gap, or a super-Earth exterior to the mini-Neptune, or both. Of those planets without mass constraints {\it Kepler}-130c is the standout case for follow-up with a minimum mass of 8.56~M$_\oplus$ in order to be consistent with photoevaporation. The system contains three detected transiting planets, the middle of which is a mini-Neptune while the other two are super-Earths with the longest period planet being the longest period super-Earth in the asteroseismic sample (the cyan points in Figure~\ref{fig:astero_multis}). Thus, with {\it Kepler}-130d a potentially long-period stripped rocky core, this system is a potentially unique laboratory for the potency of photoevaporation at long periods.     

\begin{table*}
\centering
\begin{tabular}{l l c c c c c c}
\hline
{\it Kepler}-id & KOI & Minimum  & $K_{\rm cir}$~[m/s] & Measured Mass  [M$_\oplus$] & Ref & Rocky Planet & Rocky  \\
& & Mass [M$_\oplus$] & & & & & Planet Mass  [M$_\oplus$] \\
\hline
100c & 41.01 &  No Solution & - & $<7.05$ & 3 & d & $4.94\pm0.45$ \\
130c & 282.01 & 8.56 & 1.91 &  - & - & d & $2.87\pm1.06$ \\
65c & 85.01 & 6.31 & 1.99  & $5.4\pm1.7$ & 4 & d & $4.8\pm0.51$ \\
23c & 168.01 & 4.16 & 1.14 &$60.2^{+11.4}_{-10.4}$ & 2& b & $8.12\pm1.64$ \\
23d & 168.02 & 2.09 & 0.51 &$17.6^{+13.7}_{-11.9}$ & 2 & b & $8.12\pm1.64$ \\
338b & 1930.01 & 2.64 & 0.65 &$30.6^{+24.2}_{-21.1}$ & 2 & e & $6.12\pm1.41$\\
338c & 1930.02 & 1.35 &0.27 & - & - & e & $6.12\pm1.41$ \\
338d & 1930.03 & 0.78 & 0.13& - & - & e & $6.12\pm1.41$ \\
107e & 117.01 & 1.65 & 0.40 & $8.60\pm3.60$ & 5 & c & $7.41\pm1.25$ \\
68b & 246.01 & 1.41 &0.50 & $7.65^{+1.37}_{-1.32}$ & 4 & c & $0.77\pm0.07$ \\
127c & 271.02 & 1.18 & 0.21& - & - & b & $5.32\pm1.81$ \\ 
127d & 271.01 & 0.77 &0.12 & - & - & b & $5.32\pm1.81$ \\
126d & 260.02 & 0.66 & 0.08 & - & - & c & $4.75\pm0.81$  \\
450c & 279.02 & 0.36 & 0.08 & - & - & d & $0.56\pm0.15$ \\
37d & 245.01 & 0.16 &0.04 &$<12.2$ & 3 & c & $0.31\pm0.02$ \\
10c & 72.02 & $< 0.1$ &$<0.02$ & $7.37^{+1.32}_{-1.19}$ & 1 &b & $4.38\pm0.31$\\
\hline
\end{tabular}
\caption{Predicted masses for planets in systems from the asteroseismic sample; systems are listed in descending order of predicted minimum mass. References for the measured masses: 1 - \citet{10c_mass}, 2 - \citet{Hadden2014}, 3 - \citet{Marcy2014}, 4 - \citet{65c_mass}, 5 - \citet{Bonomo2019} }\label{tab:astero}
\end{table*}

\subsubsection{Comparison to measured masses}
Of the nine planets with measured mass-constraints seven are clearly consistent, and {\it Kepler}-65c is consistent within $1\sigma$. 
However, no solution can be found for the {\it Kepler}-100c/d planets at the $>3\sigma$ level that is consistent with the photoevaporation model, this is discussed further in Section~\ref{sec:discuss_incon}. 

\subsection{CKS sample}
The CKS sample of planets by \citet{Petigura2017,Johnson2017} contains 457 multi-planet systems \citep{Weiss2018}, 190 of which contain planets that straddle the gap with their mean radii (which we fix to occur at 1.85~R$_\oplus$, {\bc independent of period}, for the CKS sample) and have mini-Neptunes with radius of $<6~$R$_\oplus$. Only 63 of these systems contain planets which straddle the gap within $2\sigma$ when accounting for radius errors, of which 2 were already analysed in the asteroseismic sample\footnote{Although the asteroseismic sample of systems is contained in the CKS sample, they do not all contain precise enough planetary radii to pass our 2$\sigma$ test.}. This leaves us 88 mini-Neptunes to perform our analysis on. The results of our analysis are shown in Table~\ref{tab:CKS}.

\subsubsection{Comparison to measured masses}
Of the 88 planets analysed in the CKS sample, 27 have some constraint on their masses. All planets are consistent with their measured masses; however the case of {\it Kepler}-172c/e system is worth noting. Only $\sim 1\%$ of our Monte-Carlo samples for {\it Kepler}-172c produce a solution, the other $\sim 99\%$ yield no solution in which both planets c and e are consistent with photoevaporation. However, all of the samples that are consistent with photoevaporation of the c/e system produce minimum masses that are consistent with the (weak) mass constraint for {\it Kepler-172c}. Having analysed $>100$ planets in total it is not too surprising there is a planet that is only consistent at the $\sim 1\%$ level. Thus, the {\it Kepler}-172 system would benefit from further observations (either in the refinement of the planetary radii and/or masses). 

Furthermore, we did analyse the {\it Kepler}-416 system\footnote{This is not included in our 88 planets discussed above.}, this contains a candidate super-Earth -- KOI1860.04 -- with a radius of $\sim 1.5$~R$_\oplus$ and period 24.8~days \citet{Johnson2017}. This would make it the longest period planet in the {\it Kepler}-416 system and as such the hardest super-Earth to completely strip. This would require the planet {\it Kepler}-416b, which has a period of $\sim 6.3$~days to have a minimum mass of $\sim 140$~M$_\oplus$ in order to be consistent with photoevaporation. However, the planet candidate KOI1860.04 was subsequently determined to have a false positive probability of 71\%\footnote{Obtained from the DR24 {\it Kepler} Reliability Report, downloaded from the NASA Exoplanet Archive on 29th August 2019}. Removing this planet candidate from the system and analysing {\it Kepler}-416b again indicates it would be consistent with the photoevaporation model provided it had a mass $\gtrsim 1$~M$_\oplus$.  

Finally, the {\it Kepler}-142c/d system yields no solution at the $>3\sigma$ level. Several of the Monte-Carlo samples do yield solutions, but with minimum masses of $\sim30-40$~M$_\oplus$ making it a nearly solid-core with $<0.1\%$ H/He envelope which also seems unlikely. Therefore, like {\it Kepler}-100, we identify the {\it Kepler}-142 as being inconsistent with the photoevaporation model, this is discussed further in section~\ref{sec:discuss_incon}.

\begin{table*}
\centering
\begin{tabular}{l l c c c c c c}
\hline
{\it Kepler}-id & KOI & Minimum  & $K_{\rm cir}$~[m/s] & Measured Mass  [M$_\oplus$] & Ref & Rocky Planet & Rocky  \\
& & Mass [M$_\oplus$] & & & & & Planet Mass  [M$_\oplus$] \\
\hline
142c & 343.01 & No Solution & - & -  & -   & d &$2.86\pm1.00$\\
176c & 520.01 & 26.56* & 8.16 &  $23.0^{+13.5}_{-8.0}$  & 1   & e &$4.01\pm1.13$\\
176d & 520.03 & 7.51 & 1.83 & $15.2^{+10.4}_{-5.8}$   & 1   & e &$4.01\pm1.13$\\
656b & 732.01 & 13.26 & 8.44& -   & -   & 732.03 &$2.62\pm0.70$\\
191d & 582.01 & 4.70 &1.82 & -   &  -  & b &$3.59\pm0.95$\\
101b & 46.01 & 4.61 & 1.77& $51.1^{+5.1}_{-4.7}$   & 2   & c &$1.94\pm0.75$\\
105b & 115.01 & 4.46 & 1.61& $5.1^{+6.3}_{-4.1}$   & 1   & c &$4.87\pm1.33$\\
226c & 749.01 & 4.16 &1.65 & $45.2^{+22.5}_{-19.1}$   & 1   & d &$3.01\pm0.81$\\
307b & 1576.01 & 3.50 &1.05 & $8.8\pm0.9$& 3   & 1576.03 &$1.79\pm0.73$\\
307c & 1576.02 & 2.53 &0.70 & $3.9\pm0.7$   & 3   & 1576.03 &$1.79\pm0.73$\\
20c & 70.01 & 2.39 &0.72 & $12.75^{+2.17}_{-2.24}$  & 4   & f &$1.08\pm0.24$\\
20d & 70.03 & 0.14 & 0.02& $10.07^{+3.97}_{-3.70}$  & 4   & f &$1.08\pm0.24$\\
- & 102.01 & 2.21 & 1.14& -  & -   & 102.02 &$1.21\pm0.37$\\
324c & 1831.01 & 1.81 &0.34 &-   & -   & 1831.03 &$2.24\pm0.64$\\
102e & 82.01 & 1.71 &0.5 & $8.93\pm2.0$  & 5   & d &$3.96\pm1.22$\\
- & 1276.01 & 1.70 & 0.41 & -   & -   & 1276.02 &$4.51\pm1.36$\\
282d & 1278.01 & 1.60 & 0.38& $61.0^{+35.9}_{-36.1}  $& 6   & c &$4.44\pm1.51$\\
282e & 1278.02 & 1.08 &  0.21&$56.2^{+16.2}_{-16.7}$  & 6   & c &$4.44\pm1.51$\\
106e & 116.02 & 0.53 & 0.1& $11.17\pm5.8$  & 5   & d &$1.16\pm0.33$\\
106c & 116.01 & 1.59 & 0.43&$10.44\pm3.2$   & 5   & d &$1.16\pm0.33$\\
189c & 574.01 & 1.47 &0.40 &$22.7^{+17.1}_{-10.6}$   & 1   & b &$3.44\pm0.89$\\
173c & 511.01 & 1.47 & 0.45& -   & -   & b &$4.61\pm1.21$\\
\hline
\end{tabular}
\caption{Those systems from the CKS sample which contain planets with predicted minimum masses $>1.4$~M$_\oplus$, systems are listed in descending order of predicted minimum mass. The full table is available online. References for the measured masses:  1 - \citet{Hadden2014}, 2 - \citet{101b_mass}, 3 - \citet{Hadden2017}, 4 - \citet{20_mass}, 5 - \citet{Marcy2014}, 6 - \citet{Xie2014},  Notes: * - 0.5\% upper-limit. The complete table is available online. }
\label{tab:CKS}
\end{table*}
\section{Discussion}
\begin{figure}
    \centering
    \includegraphics[width=\columnwidth]{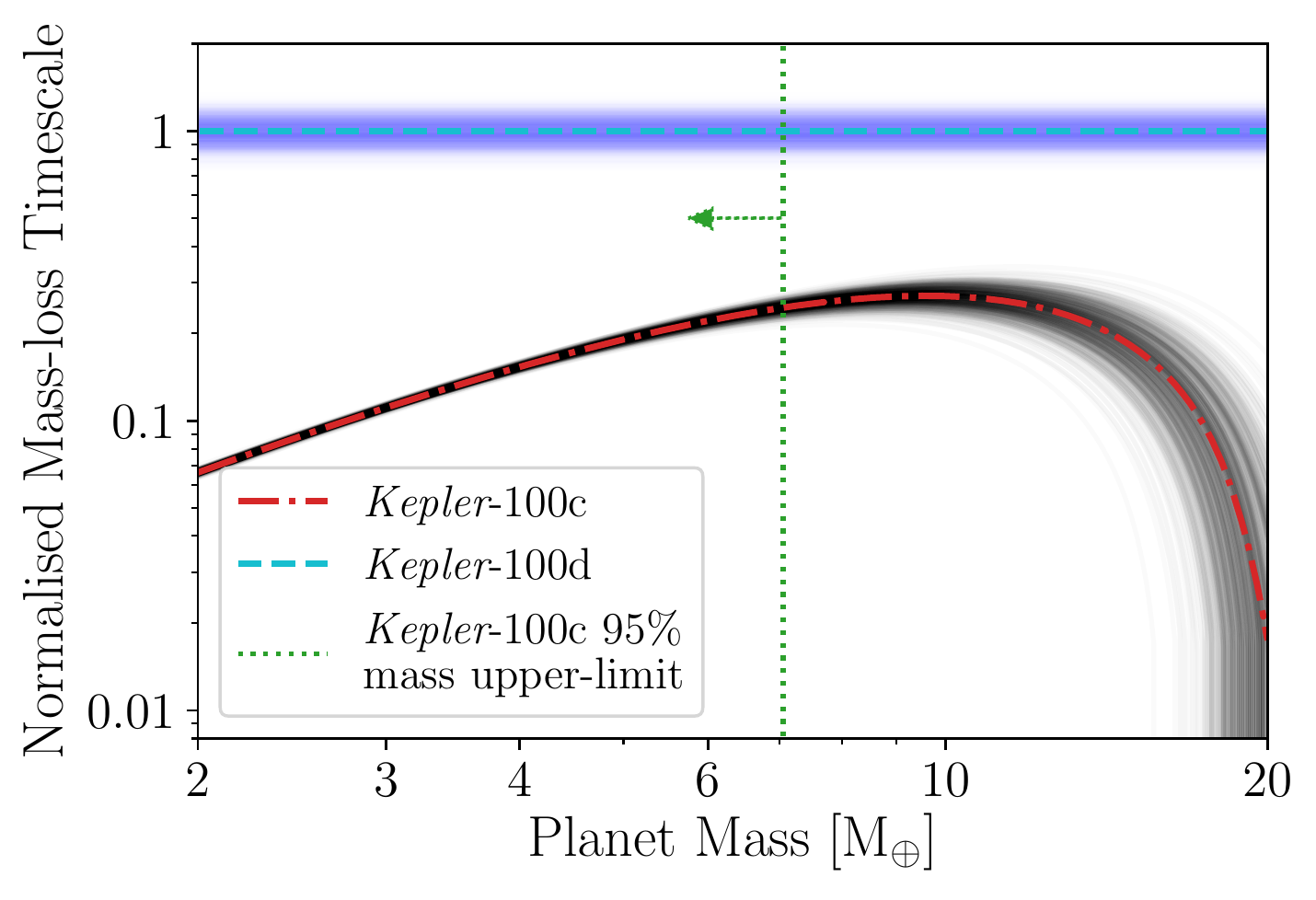}
    \caption{The mass-loss timescales  as a function of core-mass for {\it Kepler}-100c \& d (normalised to the mean value for {\it Kepler}-100d). The dot-dashed and dashed lines show the mean values and the translucent lines show 500 random draws. The dotted-line indicates the 95\% upper-mass limit for {\it Kepler}-100c \citep{Marcy2014}. This Figure is similar to the overview example shown in Figure~\ref{fig:compare}.   }
    \label{fig:kepler100}
    \centering
    \includegraphics[width=\columnwidth]{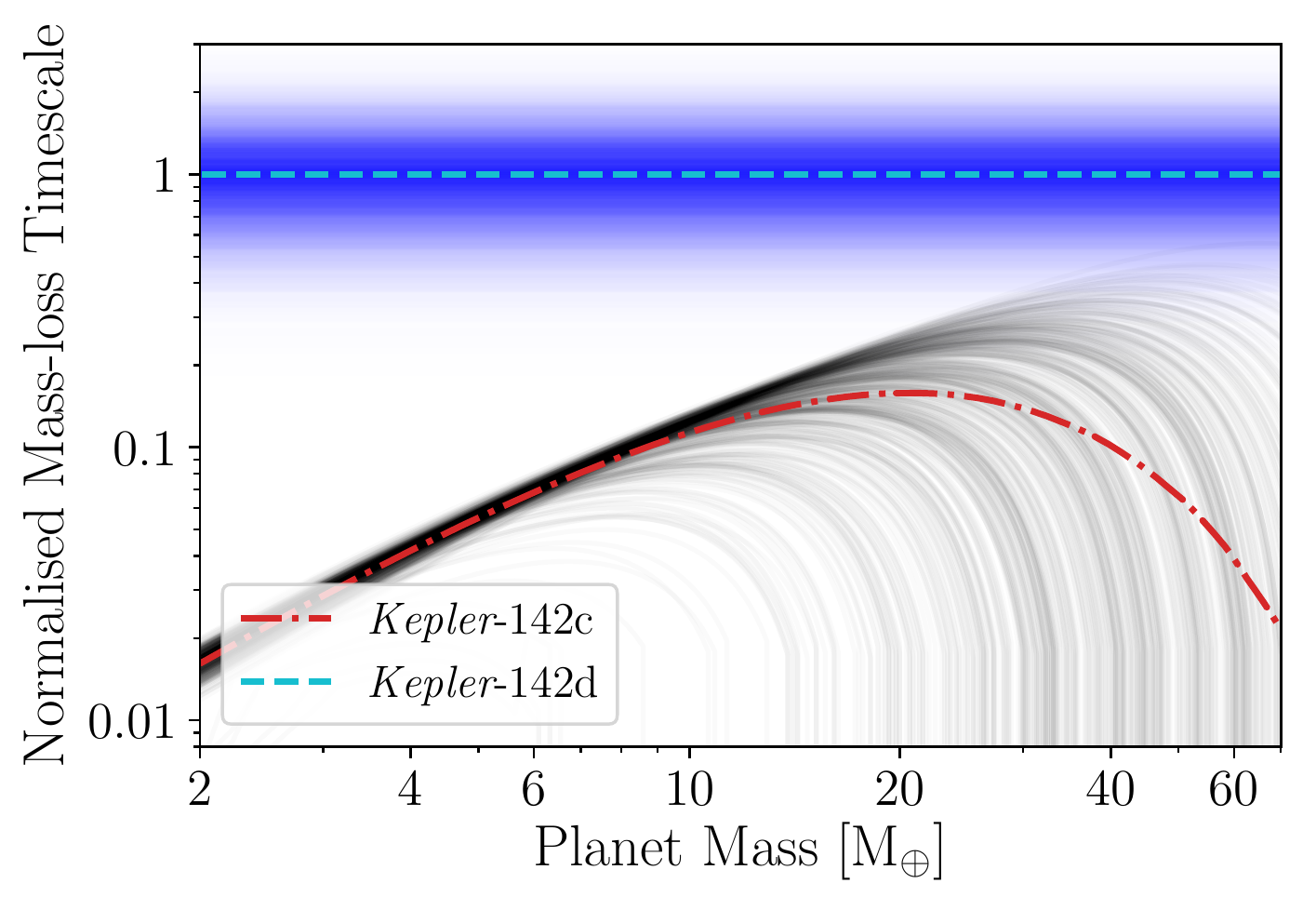}
    \caption{Same as above but for the {\it Kepler}-142c \& d system.}
    \label{fig:kepler142}
\end{figure}
Our results indicate that the photoevaporation model is in excellent agreement with the majority of observed multi-planet systems. We find 71 out of 73 systems and 35 out of the 36 planets with mass constraints are consistent with the photoevaporation model, despite 16 of those systems containing a super-Earth that is exterior to a mini-Neptune. Thus our first pass at testing the photoevaporation model with a large number of exoplanets indicates that they are consistent with the hypothesis that photoevaporation created the radius-gap, and the assumptions inferred from the photoevaporation model discussed in Section~\ref{sec:assump}. 

\subsection{Inconsistent Systems}\label{sec:discuss_incon}

However, two out of the 73 systems ({\it Kepler}-100 and {\it Kepler}-142) contained a mini-Neptune that could not retain its H/He envelope if the longest period super-Earth had lost one due to photoevaporation. These inconsistencies are demonstrated in Figure~\ref{fig:kepler100} for {\it Kepler}-100 and Figure~\ref{fig:kepler142} for {\it Kepler}-142, where we plot the mass-loss timescales for the super-Earth and mini-Neptune as a function of the mini-Neptune's core mass (i.e. a version of the right-panel of Figure~\ref{fig:compare}). These plots show that the {\it Kepler}-100 system is clearly inconsistent and that the {\it Kepler}-142 system is likely to be inconsistent (note the curves overlap, but only at high-masses ($\gtrsim 30$~M$_\oplus$), and this only occurs $<0.1$\% of the time, as discussed previously). 

\begin{figure}
    \centering
    \includegraphics[width=\columnwidth]{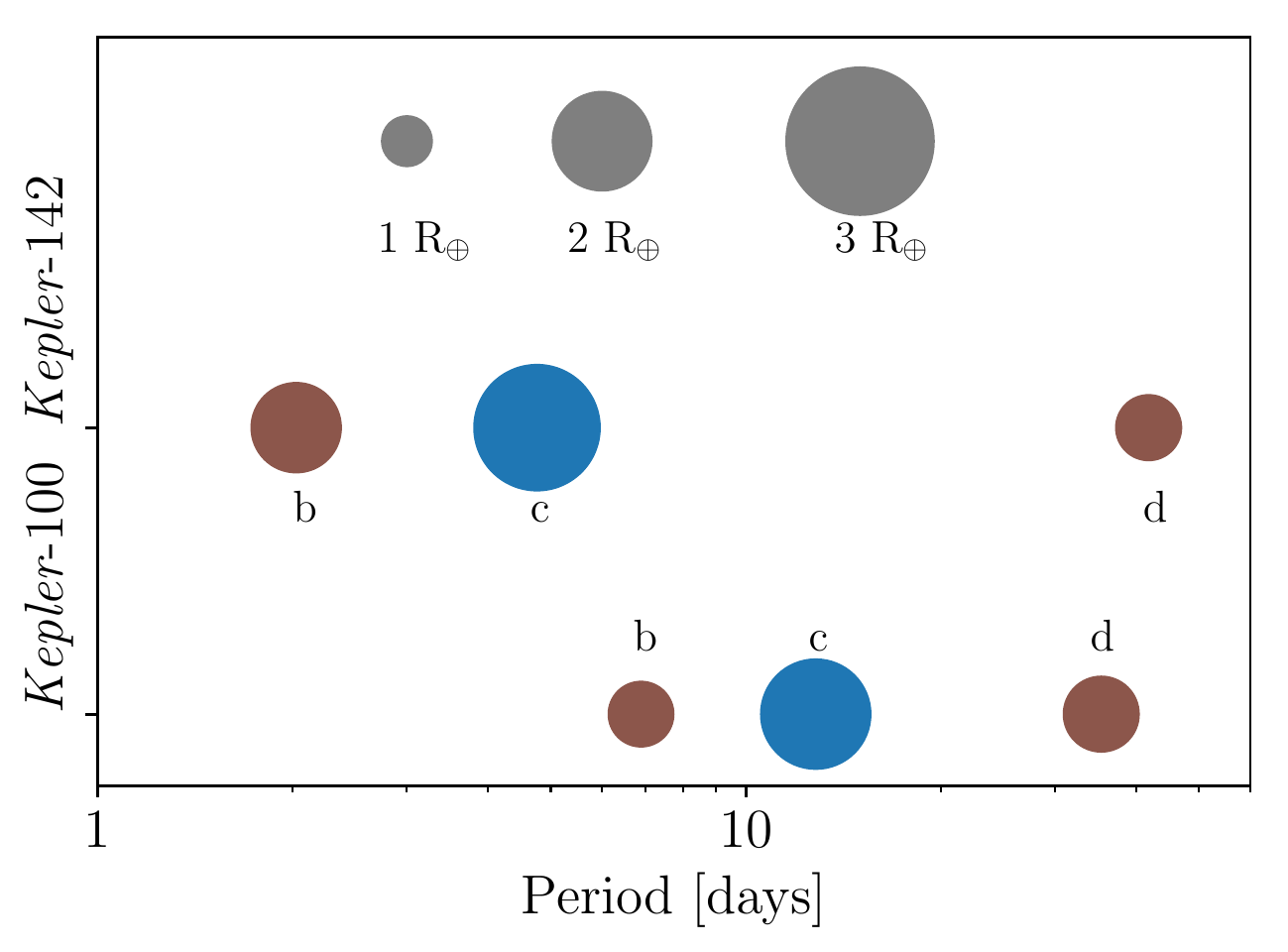}
    \caption{The architecture of {\it Kepler}-100 \& 142 systems shown in terms of the size and orbital period of their planets. Those planets identified as lying below the gap (and hence termed super-Earths in this work) are shown in brown and those identified as lying above the gap (hence termed mini-Neptunes) are shown in blue.}
    \label{fig:arch}
\end{figure}

However, what is striking is that these two systems have a fairly similar architecture, as shown in Figure~\ref{fig:arch}. Both systems contain three transiting planets, with the middle planet being above the radius gap, while the other two are below. What is also similar is that the period ratio between the interior two planets is $\sim 2$ (but not close to the 2:1 mean-motion resonance), while the period ratio between the exterior planets is much larger. Therefore, the dissimilar sizes and period ratios make these two systems rather unlike the standard {\it Kepler} multi-planet systems which have similar intra-system period and radius ratios \citep{Weiss2018,Weiss2019}. 

Experimentation with changing the representative timescale for evaporation between for a wide range of plausible (and implausible) values does not bring these systems into agreement, neither does changing the core-composition if one requires it to be identical in both planets.

{\bc A trivial solution would be that the mass-loss efficiency is wrong. To bring these systems in line would require that the efficiency is underestimated for the outer planet relative to the planet above the gap (to allow the outer planets to lose their atmospheres faster). Such a solution seems unlikely as the outer planets are fairly low mass where the mass-loss efficiencies are already high. Further, these outer planets are at quite long orbital periods, where they might be close to the point of transitioning from hydrodynamic to Jeans escape \citep{Owen2012}. Therefore, it is more likely the mass-loss efficiency for these planets is overestimated, rather than underestimated.}

However, a solution can be found if the core-composition of the mini-Neptune remains Earth-like, while the core density of the exterior super-Earth is lowered (making it easier to strip). Only a small increase in the core density (from 1/3 iron to $\lesssim$ 5\% iron) for {\it Kepler}-142d is required to make the system consistent. Though, for the {\it Kepler}-100 system, planet d requires a core-composition that is $\gtrsim 25\%$ ice/water. Such dissimilar core densities are proposed in some formation scenarios \citep[e.g.][]{Raymond2018}. There is some evidence that this could be the solution, as the measured mass-limit (95\% upper-limit) for {\it Kepler}-100d is 3~M$_\oplus$ \citep{Marcy2014}, significantly smaller than the assumed mass in our analysis, which was based on its radius and adopting an Earth-like composition of $4.95\pm0.45$~M$_\oplus$ (see Table~\ref{tab:astero}). To satisfy the mass upper-limit if the planet contains no H/He atmosphere requires {\it Kepler}-100d to contain some water or ice. Therefore, it is plausible that the outer planets in these two systems having water/ice rich compositions can provide the solution to why they appear inconsistent with the photoevaporation model. We do however caution that the mass-limits of {\it Kepler}-100d assume a circular orbit \citep{Marcy2014}; however, it is now known from the asteroseismic analysis that it has a significant eccentricity of $0.38^{0.12}_{-0.16}$ \citep{VanEylen2015}. Therefore it still remains open whether {\it Kepler}-100d could be ice/water rich or not.

The fact that {\it Kepler}-100d is eccentric and at a significantly larger period ratio could point to another possible scenario: a giant-impact occurred after disc dispersal \citep[e.g.][]{Inamdar2016}. Giant-impacts are efficient at removing H/He atmospheres, as well as enhancing the affect of photoevaporation \citep[e.g.][]{Biersteker2018}.  There appears to be no eccentricity or mass measurement available for {\it Kepler}-142d; however, the similar orbital architecture indicates that this is also a plausible scenario for the {\it Kepler}-142 system. 

Alternatively, the outer planets could have formed after disc dispersal in these two systems, never accreting an initial H/He envelope or these two systems are indicating the photoevaporative efficiency scaling is incorrect, meaning these two planets still have H/He atmospheres. 

\subsection{Future comparisons}
We have identified two systems that are inconsistent with photoevaporation ({\it Kepler}-100 \& 142); however, further follow-up of these systems would illuminate as to the origin of this disagreement: dissimilar core densities, giant-impacts or something else. For {\it Kepler}-100, refitting the masses assuming a non-circular orbit is a priority. For {\it Kepler}-142, using the precise constraints on the stellar density from the CKS survey and {\it GAIA} \citep[e.g.][]{Petigura2017,Fulton2018} to refit the transit photometry would improve radii and eccentricity estimates, as well as obtaining follow-up RV masses. {\bc Further, observing the system with {\it CHEOPS} to obtain precise radii estimates would also improve the constraints}. {\it Kepler}-176 is also a system worth investigating in greater detail: as well as refitting the transit photometry {\bc or observing the system with {\it CHEOPS}}, improving the transit-timing mass-measurements once the {\it Kepler} field is observed by {\it TESS} will indicate whether this system is consistent with the photoevaporation model or not. {\bc Further, we note the predicted masses of {\it Kepler}-176c \& d imply they are separated by $\sim 10$ mutual hill radii, implying they may be close to dynamical stability.}

Furthermore, we have provided a list of 12 planets which have estimated RV semi-amplitudes $>1$~m/s which should be prime targets for spectroscopic follow-up observations. Our method should be directly applicable to multi-planet systems that {\it TESS} finds and where RV follow-up is easier. In anticipation of such possibilities the software used to estimate the minimum masses is available to the public.  We also note that using RV follow-up to test the photoevaporation model may require going beyond $3\sigma$ mass measurements, and that upper-limits can be equally useful\footnote{We suspect there may even be RV data already available to test some of the {\it Kepler}-systems in Tables~\ref{tab:astero} \& \ref{tab:CKS} that is yet to be published as a 3$\sigma$ mass detection.} .

\section{Summary}
We have shown how the photoevaporation model can be used to estimate the minimum masses of planets that are hypothesised to posses large hydrogen/helium atmospheres if they reside in a system that also contains a planet below the exoplanet radius-gap. Our approach is valuable as it is independent of the host star's uncertain EUV/X-ray history. This method essentially answers the following question: given photoevaporation had to completely strip the planet that resides below the radius-gap, how massive does the planet above the radius-gap need to be to retain its hydrogen/helium envelope? Unsurprisingly, the most stringent constraints come from systems in which the planet below the gap is large, and is exterior to the planet above the gap.

We have applied this method to 104 exoplanets that likely reside above the radius gap (within 2$\sigma$) in 73 systems that likely contain a planet that resides below the radius-gap (within 2$\sigma$). In general, we find excellent agreement between the photoevaporation model and those planets with measured mass constraints, indicating that photoevaporation is the cause of the radius-gap. Only two planets ({\it Kepler} 100c and 142c) are inconsistent with the photoevaporation at the $>3\sigma$ level; and we speculate that these systems could either contain planets with dissimilar core compositions, underwent a giant impact, had planets that both formed before and after the gas disc dispersed, {\bc or simply the mass-loss rates are incorrect}. Any of these hypothesis make these systems unusual in terms of their evolutionary pathway when compared to the bulk of the population, and warrant more detailed study. 

We have identified 12 planets with RV semi-amplitudes $\gtrsim 1$ m~s$^{-1}$, the vast majority of which only have weak TTV mass constraints or no measured mass. Therefore, these planets would be prime targets for testing the photoevaporation model either with RV follow-up or additional studies of TTVs when {\it TESS} observes the {\it Kepler} field. 

Finally, our method can be applied to any multi-planet systems that contain planets which straddle the radius-gap. Therefore, ongoing and future transit surveys should provide a detailed test of the photoevaporation model, provided they are combined with a well designed follow-up program to measure masses. We emphasis that upper-mass limits are equally useful as mass measurements, since our method predicts a \emph{minimum} planet mass. {\bc We also suspect this method of comparing planets in the same system could be applied to other mass-loss mechanisms e.g. core-powered mass-loss \citep{Ginzburg2018}}.

\section*{Acknowledgements}

We thank the referee for a report which improved the manuscript. JEO is supported by a Royal Society University Research Fellowship. We are grateful to Vincent van Eylen, the CKS team and the {\it Kepler} team for making their work publicly available to allow us to perform this analysis. This research has made use of the NASA Exoplanet Archive, which is operated by the California Institute of Technology, under contract with the National Aeronautics and Space Administration under the Exoplanet Exploration Program. This work was performed using the DiRAC Data Intensive service at Leicester, operated by the University of Leicester IT Services, which forms part of the STFC DiRAC HPC Facility (www.dirac.ac.uk). The equipment was funded by BEIS capital funding via STFC capital grants ST/K000373/1 and ST/R002363/1 and STFC DiRAC Operations grant ST/R001014/1. DiRAC is part of the National e-Infrastructure.





\bibliographystyle{mnras}
\bibliography{bib_paper}

\begin{thebibliography}{}
\makeatletter
\relax
\def\mn@urlcharsother{\let\do\@makeother \do\$\do\&\do\#\do\^\do\_\do\%\do\~}
\def\mn@doi{\begingroup\mn@urlcharsother \@ifnextchar [ {\mn@doi@}
  {\mn@doi@[]}}
\def\mn@doi@[#1]#2{\def\@tempa{#1}\ifx\@tempa\@empty \href
  {http://dx.doi.org/#2} {doi:#2}\else \href {http://dx.doi.org/#2} {#1}\fi
  \endgroup}
\def\mn@eprint#1#2{\mn@eprint@#1:#2::\@nil}
\def\mn@eprint@arXiv#1{\href {http://arxiv.org/abs/#1} {{\tt arXiv:#1}}}
\def\mn@eprint@dblp#1{\href {http://dblp.uni-trier.de/rec/bibtex/#1.xml}
  {dblp:#1}}
\def\mn@eprint@#1:#2:#3:#4\@nil{\def\@tempa {#1}\def\@tempb {#2}\def\@tempc
  {#3}\ifx \@tempc \@empty \let \@tempc \@tempb \let \@tempb \@tempa \fi \ifx
  \@tempb \@empty \def\@tempb {arXiv}\fi \@ifundefined
  {mn@eprint@\@tempb}{\@tempb:\@tempc}{\expandafter \expandafter \csname
  mn@eprint@\@tempb\endcsname \expandafter{\@tempc}}}

\bibitem[\protect\citeauthoryear{{Allart} et~al.,}{{Allart}
  et~al.}{2018}]{Allart2018}
{Allart} R.,  et~al., 2018, \mn@doi [Science] {10.1126/science.aat5879}, \href
  {https://ui.adsabs.harvard.edu/abs/2018Sci...362.1384A} {362, 1384}

\bibitem[\protect\citeauthoryear{{Baraffe}, {Chabrier}, {Barman}, {Selsis},
  {Allard}  \& {Hauschildt}}{{Baraffe} et~al.}{2005}]{Baraffe2005}
{Baraffe} I.,  {Chabrier} G.,  {Barman} T.~S.,  {Selsis} F.,  {Allard} F.,
  {Hauschildt} P.~H.,  2005, \mn@doi [\aap] {10.1051/0004-6361:200500123},
  \href {https://ui.adsabs.harvard.edu/abs/2005A&A...436L..47B} {436, L47}

\bibitem[\protect\citeauthoryear{{Barclay}, {Pepper}  \& {Quintana}}{{Barclay}
  et~al.}{2018}]{Barclay2018}
{Barclay} T.,  {Pepper} J.,   {Quintana} E.~V.,  2018, \mn@doi [\apjs]
  {10.3847/1538-4365/aae3e9}, \href
  {https://ui.adsabs.harvard.edu/abs/2018ApJS..239....2B} {239, 2}

\bibitem[\protect\citeauthoryear{{Biersteker} \& {Schlichting}}{{Biersteker} \&
  {Schlichting}}{2019}]{Biersteker2018}
{Biersteker} J.~B.,  {Schlichting} H.~E.,  2019, \mn@doi [\mnras]
  {10.1093/mnras/stz738}, \href
  {https://ui.adsabs.harvard.edu/abs/2019MNRAS.485.4454B} {485, 4454}

\bibitem[\protect\citeauthoryear{{Bonomo} et~al.,}{{Bonomo}
  et~al.}{2014}]{101b_mass}
{Bonomo} A.~S.,  et~al., 2014, \mn@doi [\aap] {10.1051/0004-6361/201424617},
  \href {https://ui.adsabs.harvard.edu/abs/2014A&A...572A...2B} {572, A2}

\bibitem[\protect\citeauthoryear{{Bonomo} et~al.,}{{Bonomo}
  et~al.}{2019}]{Bonomo2019}
{Bonomo} A.~S.,  et~al., 2019, \mn@doi [Nature Astronomy]
  {10.1038/s41550-018-0684-9}, \href
  {https://ui.adsabs.harvard.edu/abs/2019NatAs...3..416B} {3, 416}

\bibitem[\protect\citeauthoryear{{Borucki} et~al.,}{{Borucki}
  et~al.}{2011}]{Borucki2011}
{Borucki} W.~J.,  et~al., 2011, \mn@doi [\apj] {10.1088/0004-637X/736/1/19},
  \href {https://ui.adsabs.harvard.edu/abs/2011ApJ...736...19B} {736, 19}

\bibitem[\protect\citeauthoryear{{Buchhave} et~al.,}{{Buchhave}
  et~al.}{2016}]{20_mass}
{Buchhave} L.~A.,  et~al., 2016, \mn@doi [\aj] {10.3847/0004-6256/152/6/160},
  \href {https://ui.adsabs.harvard.edu/abs/2016AJ....152..160B} {152, 160}

\bibitem[\protect\citeauthoryear{{Carter} et~al.,}{{Carter}
  et~al.}{2012}]{Carter2012}
{Carter} J.~A.,  et~al., 2012, \mn@doi [Science] {10.1126/science.1223269},
  \href {https://ui.adsabs.harvard.edu/abs/2012Sci...337..556C} {337, 556}

\bibitem[\protect\citeauthoryear{{Chen} \& {Rogers}}{{Chen} \&
  {Rogers}}{2016}]{Chen2016}
{Chen} H.,  {Rogers} L.~A.,  2016, \mn@doi [\apj]
  {10.3847/0004-637X/831/2/180}, \href
  {https://ui.adsabs.harvard.edu/abs/2016ApJ...831..180C} {831, 180}

\bibitem[\protect\citeauthoryear{{Dorn}, {Harrison}, {Bonsor}  \&
  {Hands}}{{Dorn} et~al.}{2019}]{Dorn2019}
{Dorn} C.,  {Harrison} J.~H.~D.,  {Bonsor} A.,   {Hands} T.~O.,  2019, \mn@doi
  [\mnras] {10.1093/mnras/sty3435}, \href
  {https://ui.adsabs.harvard.edu/abs/2019MNRAS.484..712D} {484, 712}

\bibitem[\protect\citeauthoryear{{Dressing} et~al.,}{{Dressing}
  et~al.}{2015}]{Dressing2015}
{Dressing} C.~D.,  et~al., 2015, \mn@doi [\apj] {10.1088/0004-637X/800/2/135},
  \href {https://ui.adsabs.harvard.edu/abs/2015ApJ...800..135D} {800, 135}

\bibitem[\protect\citeauthoryear{{Ehrenreich} et~al.,}{{Ehrenreich}
  et~al.}{2015}]{Ehrenreich2015}
{Ehrenreich} D.,  et~al., 2015, \mn@doi [\nat] {10.1038/nature14501}, \href
  {https://ui.adsabs.harvard.edu/abs/2015Natur.522..459E} {522, 459}

\bibitem[\protect\citeauthoryear{{Fortney}, {Marley}  \& {Barnes}}{{Fortney}
  et~al.}{2007}]{Fortney2007}
{Fortney} J.~J.,  {Marley} M.~S.,   {Barnes} J.~W.,  2007, \mn@doi [\apj]
  {10.1086/512120}, \href
  {https://ui.adsabs.harvard.edu/abs/2007ApJ...659.1661F} {659, 1661}

\bibitem[\protect\citeauthoryear{{Fressin} et~al.,}{{Fressin}
  et~al.}{2013}]{Fressin2013}
{Fressin} F.,  et~al., 2013, \mn@doi [\apj] {10.1088/0004-637X/766/2/81}, \href
  {https://ui.adsabs.harvard.edu/abs/2013ApJ...766...81F} {766, 81}

\bibitem[\protect\citeauthoryear{{Fulton} \& {Petigura}}{{Fulton} \&
  {Petigura}}{2018}]{Fulton2018}
{Fulton} B.~J.,  {Petigura} E.~A.,  2018, \mn@doi [\aj]
  {10.3847/1538-3881/aae828}, \href
  {https://ui.adsabs.harvard.edu/abs/2018AJ....156..264F} {156, 264}

\bibitem[\protect\citeauthoryear{{Fulton} et~al.,}{{Fulton}
  et~al.}{2017}]{Fulton2017}
{Fulton} B.~J.,  et~al., 2017, \mn@doi [\aj] {10.3847/1538-3881/aa80eb}, \href
  {https://ui.adsabs.harvard.edu/abs/2017AJ....154..109F} {154, 109}

\bibitem[\protect\citeauthoryear{{Ginzburg}, {Schlichting}  \&
  {Sari}}{{Ginzburg} et~al.}{2018}]{Ginzburg2018}
{Ginzburg} S.,  {Schlichting} H.~E.,   {Sari} R.,  2018, \mn@doi [\mnras]
  {10.1093/mnras/sty290}, \href
  {https://ui.adsabs.harvard.edu/abs/2018MNRAS.476..759G} {476, 759}

\bibitem[\protect\citeauthoryear{{G{\"u}nther}, {Queloz}, {Demory}  \&
  {Bouchy}}{{G{\"u}nther} et~al.}{2017}]{Gunther2017}
{G{\"u}nther} M.~N.,  {Queloz} D.,  {Demory} B.-O.,   {Bouchy} F.,  2017,
  \mn@doi [\mnras] {10.1093/mnras/stw2908}, \href
  {https://ui.adsabs.harvard.edu/abs/2017MNRAS.465.3379G} {465, 3379}

\bibitem[\protect\citeauthoryear{{Gupta} \& {Schlichting}}{{Gupta} \&
  {Schlichting}}{2019a}]{Gupta2019b}
{Gupta} A.,  {Schlichting} H.~E.,  2019a, arXiv e-prints, \href
  {https://ui.adsabs.harvard.edu/abs/2019arXiv190703732G} {p. arXiv:1907.03732}

\bibitem[\protect\citeauthoryear{{Gupta} \& {Schlichting}}{{Gupta} \&
  {Schlichting}}{2019b}]{Gupta2019}
{Gupta} A.,  {Schlichting} H.~E.,  2019b, \mn@doi [\mnras]
  {10.1093/mnras/stz1230}, \href
  {https://ui.adsabs.harvard.edu/abs/2019MNRAS.487...24G} {487, 24}

\bibitem[\protect\citeauthoryear{{Hadden} \& {Lithwick}}{{Hadden} \&
  {Lithwick}}{2014}]{Hadden2014}
{Hadden} S.,  {Lithwick} Y.,  2014, \mn@doi [\apj]
  {10.1088/0004-637X/787/1/80}, \href
  {https://ui.adsabs.harvard.edu/abs/2014ApJ...787...80H} {787, 80}

\bibitem[\protect\citeauthoryear{{Hadden} \& {Lithwick}}{{Hadden} \&
  {Lithwick}}{2017}]{Hadden2017}
{Hadden} S.,  {Lithwick} Y.,  2017, \mn@doi [\aj] {10.3847/1538-3881/aa71ef},
  \href {https://ui.adsabs.harvard.edu/abs/2017AJ....154....5H} {154, 5}

\bibitem[\protect\citeauthoryear{{Howard} et~al.,}{{Howard}
  et~al.}{2010}]{Howard2010}
{Howard} A.~W.,  et~al., 2010, \mn@doi [Science] {10.1126/science.1194854},
  \href {https://ui.adsabs.harvard.edu/abs/2010Sci...330..653H} {330, 653}

\bibitem[\protect\citeauthoryear{{Inamdar} \& {Schlichting}}{{Inamdar} \&
  {Schlichting}}{2016}]{Inamdar2016}
{Inamdar} N.~K.,  {Schlichting} H.~E.,  2016, \mn@doi [\apjl]
  {10.3847/2041-8205/817/2/L13}, \href
  {https://ui.adsabs.harvard.edu/abs/2016ApJ...817L..13I} {817, L13}

\bibitem[\protect\citeauthoryear{{Jankovic}, {Owen}  \& {Mohanty}}{{Jankovic}
  et~al.}{2019}]{Jankovic2019}
{Jankovic} M.~R.,  {Owen} J.~E.,   {Mohanty} S.,  2019, \mn@doi [\mnras]
  {10.1093/mnras/stz004}, \href
  {https://ui.adsabs.harvard.edu/abs/2019MNRAS.484.2296J} {484, 2296}

\bibitem[\protect\citeauthoryear{{Jin} \& {Mordasini}}{{Jin} \&
  {Mordasini}}{2018}]{Jin2018}
{Jin} S.,  {Mordasini} C.,  2018, \mn@doi [\apj] {10.3847/1538-4357/aa9f1e},
  \href {https://ui.adsabs.harvard.edu/abs/2018ApJ...853..163J} {853, 163}

\bibitem[\protect\citeauthoryear{{Jin}, {Mordasini}, {Parmentier}, {van
  Boekel}, {Henning}  \& {Ji}}{{Jin} et~al.}{2014}]{Jin2014}
{Jin} S.,  {Mordasini} C.,  {Parmentier} V.,  {van Boekel} R.,  {Henning} T.,
  {Ji} J.,  2014, \mn@doi [\apj] {10.1088/0004-637X/795/1/65}, \href
  {https://ui.adsabs.harvard.edu/abs/2014ApJ...795...65J} {795, 65}

\bibitem[\protect\citeauthoryear{{Johnson} et~al.,}{{Johnson}
  et~al.}{2017}]{Johnson2017}
{Johnson} J.~A.,  et~al., 2017, \mn@doi [The Astronomical Journal]
  {10.3847/1538-3881/aa80e7}, \href
  {https://ui.adsabs.harvard.edu/abs/2017AJ....154..108J} {154, 108}

\bibitem[\protect\citeauthoryear{{Jontof-Hutter} et~al.,}{{Jontof-Hutter}
  et~al.}{2016}]{JontofHutter2016}
{Jontof-Hutter} D.,  et~al., 2016, \mn@doi [\apj] {10.3847/0004-637X/820/1/39},
  \href {https://ui.adsabs.harvard.edu/abs/2016ApJ...820...39J} {820, 39}

\bibitem[\protect\citeauthoryear{{Kubyshkina} et~al.,}{{Kubyshkina}
  et~al.}{2018}]{Kubyshkina2018}
{Kubyshkina} D.,  et~al., 2018, \mn@doi [\aap] {10.1051/0004-6361/201833737},
  \href {https://ui.adsabs.harvard.edu/abs/2018A&A...619A.151K} {619, A151}

\bibitem[\protect\citeauthoryear{{Lammer}, {Selsis}, {Ribas}, {Guinan}, {Bauer}
   \& {Weiss}}{{Lammer} et~al.}{2003}]{Lammer2003}
{Lammer} H.,  {Selsis} F.,  {Ribas} I.,  {Guinan} E.~F.,  {Bauer} S.~J.,
  {Weiss} W.~W.,  2003, \mn@doi [\apjl] {10.1086/380815}, \href
  {https://ui.adsabs.harvard.edu/abs/2003ApJ...598L.121L} {598, L121}

\bibitem[\protect\citeauthoryear{{Lecavelier Des Etangs} et~al.,}{{Lecavelier
  Des Etangs} et~al.}{2010}]{Lecavelier2010}
{Lecavelier Des Etangs} A.,  et~al., 2010, \mn@doi [\aap]
  {10.1051/0004-6361/200913347}, \href
  {https://ui.adsabs.harvard.edu/abs/2010A&A...514A..72L} {514, A72}

\bibitem[\protect\citeauthoryear{{Lopez} \& {Fortney}}{{Lopez} \&
  {Fortney}}{2013}]{Lopez2013}
{Lopez} E.~D.,  {Fortney} J.~J.,  2013, \mn@doi [\apj]
  {10.1088/0004-637X/776/1/2}, \href
  {https://ui.adsabs.harvard.edu/abs/2013ApJ...776....2L} {776, 2}

\bibitem[\protect\citeauthoryear{{Lopez}, {Fortney}  \& {Miller}}{{Lopez}
  et~al.}{2012}]{Lopez2012}
{Lopez} E.~D.,  {Fortney} J.~J.,   {Miller} N.,  2012, \mn@doi [\apj]
  {10.1088/0004-637X/761/1/59}, \href
  {https://ui.adsabs.harvard.edu/abs/2012ApJ...761...59L} {761, 59}

\bibitem[\protect\citeauthoryear{{Marcy} et~al.,}{{Marcy}
  et~al.}{2014}]{Marcy2014}
{Marcy} G.~W.,  et~al., 2014, \mn@doi [\apjs] {10.1088/0067-0049/210/2/20},
  \href {https://ui.adsabs.harvard.edu/abs/2014ApJS..210...20M} {210, 20}

\bibitem[\protect\citeauthoryear{{Mayor} et~al.,}{{Mayor}
  et~al.}{2011}]{Mayor2011}
{Mayor} M.,  et~al., 2011, arXiv e-prints, \href
  {https://ui.adsabs.harvard.edu/abs/2011arXiv1109.2497M} {p. arXiv:1109.2497}

\bibitem[\protect\citeauthoryear{{Mills} et~al.,}{{Mills}
  et~al.}{2019}]{65c_mass}
{Mills} S.~M.,  et~al., 2019, \mn@doi [The Astronomical Journal]
  {10.3847/1538-3881/ab0899}, \href
  {https://ui.adsabs.harvard.edu/abs/2019AJ....157..145M} {157, 145}

\bibitem[\protect\citeauthoryear{{Mulders}, {Pascucci}, {Apai}  \&
  {Ciesla}}{{Mulders} et~al.}{2018}]{Mulders2018}
{Mulders} G.~D.,  {Pascucci} I.,  {Apai} D.,   {Ciesla} F.~J.,  2018, \mn@doi
  [\aj] {10.3847/1538-3881/aac5ea}, \href
  {https://ui.adsabs.harvard.edu/abs/2018AJ....156...24M} {156, 24}

\bibitem[\protect\citeauthoryear{{Murray-Clay}, {Chiang}  \&
  {Murray}}{{Murray-Clay} et~al.}{2009}]{MurrayClay2009}
{Murray-Clay} R.~A.,  {Chiang} E.~I.,   {Murray} N.,  2009, \mn@doi [\apj]
  {10.1088/0004-637X/693/1/23}, \href
  {https://ui.adsabs.harvard.edu/abs/2009ApJ...693...23M} {693, 23}

\bibitem[\protect\citeauthoryear{{Owen}}{{Owen}}{2019}]{Owen2019}
{Owen} J.~E.,  2019, \mn@doi [Annual Review of Earth and Planetary Sciences]
  {10.1146/annurev-earth-053018-060246}, \href
  {https://ui.adsabs.harvard.edu/abs/2019AREPS..47...67O} {47, 67}

\bibitem[\protect\citeauthoryear{{Owen} \& {Adams}}{{Owen} \&
  {Adams}}{2019}]{OA19}
{Owen} J.~E.,  {Adams} F.~C.,  2019, \mnras\, submitted, \href
  {https://ui.adsabs.harvard.edu/abs/2014MNRAS.444.3761O} {}

\bibitem[\protect\citeauthoryear{{Owen} \& {Jackson}}{{Owen} \&
  {Jackson}}{2012}]{Owen2012}
{Owen} J.~E.,  {Jackson} A.~P.,  2012, \mn@doi [\mnras]
  {10.1111/j.1365-2966.2012.21481.x}, \href
  {https://ui.adsabs.harvard.edu/abs/2012MNRAS.425.2931O} {425, 2931}

\bibitem[\protect\citeauthoryear{{Owen} \& {Morton}}{{Owen} \&
  {Morton}}{2016}]{Owen2016}
{Owen} J.~E.,  {Morton} T.~D.,  2016, \mn@doi [\apjl]
  {10.3847/2041-8205/819/1/L10}, \href
  {https://ui.adsabs.harvard.edu/abs/2016ApJ...819L..10O} {819, L10}

\bibitem[\protect\citeauthoryear{{Owen} \& {Wu}}{{Owen} \&
  {Wu}}{2013}]{Owen2013}
{Owen} J.~E.,  {Wu} Y.,  2013, \mn@doi [\apj] {10.1088/0004-637X/775/2/105},
  \href {https://ui.adsabs.harvard.edu/abs/2013ApJ...775..105O} {775, 105}

\bibitem[\protect\citeauthoryear{{Owen} \& {Wu}}{{Owen} \&
  {Wu}}{2017}]{Owen2017}
{Owen} J.~E.,  {Wu} Y.,  2017, \mn@doi [\apj] {10.3847/1538-4357/aa890a}, \href
  {https://ui.adsabs.harvard.edu/abs/2017ApJ...847...29O} {847, 29}

\bibitem[\protect\citeauthoryear{{Paxton}, {Bildsten}, {Dotter}, {Herwig},
  {Lesaffre}  \& {Timmes}}{{Paxton} et~al.}{2011}]{Paxton2011}
{Paxton} B.,  {Bildsten} L.,  {Dotter} A.,  {Herwig} F.,  {Lesaffre} P.,
  {Timmes} F.,  2011, \mn@doi [\apjs] {10.1088/0067-0049/192/1/3}, \href
  {https://ui.adsabs.harvard.edu/abs/2011ApJS..192....3P} {192, 3}

\bibitem[\protect\citeauthoryear{{Paxton} et~al.,}{{Paxton}
  et~al.}{2013}]{Paxton2013}
{Paxton} B.,  et~al., 2013, \mn@doi [\apjs] {10.1088/0067-0049/208/1/4}, \href
  {https://ui.adsabs.harvard.edu/abs/2013ApJS..208....4P} {208, 4}

\bibitem[\protect\citeauthoryear{{Petigura} et~al.,}{{Petigura}
  et~al.}{2017}]{Petigura2017}
{Petigura} E.~A.,  et~al., 2017, \mn@doi [The Astronomical Journal]
  {10.3847/1538-3881/aa80de}, \href
  {https://ui.adsabs.harvard.edu/abs/2017AJ....154..107P} {154, 107}

\bibitem[\protect\citeauthoryear{{Rajpaul}, {Buchhave}  \& {Aigrain}}{{Rajpaul}
  et~al.}{2017}]{10c_mass}
{Rajpaul} V.,  {Buchhave} L.~A.,   {Aigrain} S.,  2017, \mn@doi [Monthly
  Notices of the Royal Astronomical Society] {10.1093/mnrasl/slx116}, \href
  {https://ui.adsabs.harvard.edu/abs/2017MNRAS.471L.125R} {471, L125}

\bibitem[\protect\citeauthoryear{{Raymond}, {Boulet}, {Izidoro}, {Esteves}  \&
  {Bitsch}}{{Raymond} et~al.}{2018}]{Raymond2018}
{Raymond} S.~N.,  {Boulet} T.,  {Izidoro} A.,  {Esteves} L.,   {Bitsch} B.,
  2018, \mn@doi [\mnras] {10.1093/mnrasl/sly100}, \href
  {https://ui.adsabs.harvard.edu/abs/2018MNRAS.479L..81R} {479, L81}

\bibitem[\protect\citeauthoryear{{Rogers}}{{Rogers}}{2015}]{Rogers2015}
{Rogers} L.~A.,  2015, \mn@doi [\apj] {10.1088/0004-637X/801/1/41}, \href
  {https://ui.adsabs.harvard.edu/abs/2015ApJ...801...41R} {801, 41}

\bibitem[\protect\citeauthoryear{{Rogers} \& {Seager}}{{Rogers} \&
  {Seager}}{2010}]{Rogers2010}
{Rogers} L.~A.,  {Seager} S.,  2010, \mn@doi [\apj]
  {10.1088/0004-637X/712/2/974}, \href
  {https://ui.adsabs.harvard.edu/abs/2010ApJ...712..974R} {712, 974}

\bibitem[\protect\citeauthoryear{{Silburt}, {Gaidos}  \& {Wu}}{{Silburt}
  et~al.}{2015}]{Silburt2015}
{Silburt} A.,  {Gaidos} E.,   {Wu} Y.,  2015, \mn@doi [\apj]
  {10.1088/0004-637X/799/2/180}, \href
  {https://ui.adsabs.harvard.edu/abs/2015ApJ...799..180S} {799, 180}

\bibitem[\protect\citeauthoryear{{Spake} et~al.,}{{Spake}
  et~al.}{2018}]{Spake2018}
{Spake} J.~J.,  et~al., 2018, \mn@doi [\nat] {10.1038/s41586-018-0067-5}, \href
  {https://ui.adsabs.harvard.edu/abs/2018Natur.557...68S} {557, 68}

\bibitem[\protect\citeauthoryear{{Thompson} et~al.,}{{Thompson}
  et~al.}{2018}]{Thompson2018}
{Thompson} S.~E.,  et~al., 2018, \mn@doi [\apjs] {10.3847/1538-4365/aab4f9},
  \href {https://ui.adsabs.harvard.edu/abs/2018ApJS..235...38T} {235, 38}

\bibitem[\protect\citeauthoryear{{Tu}, {Johnstone}, {G{\"u}del}  \&
  {Lammer}}{{Tu} et~al.}{2015}]{Tu2015}
{Tu} L.,  {Johnstone} C.~P.,  {G{\"u}del} M.,   {Lammer} H.,  2015, \mn@doi
  [\aap] {10.1051/0004-6361/201526146}, \href
  {https://ui.adsabs.harvard.edu/abs/2015A&A...577L...3T} {577, L3}

\bibitem[\protect\citeauthoryear{{Valencia}, {Sasselov}  \&
  {O'Connell}}{{Valencia} et~al.}{2007}]{Valencia2007}
{Valencia} D.,  {Sasselov} D.~D.,   {O'Connell} R.~J.,  2007, \mn@doi [\apj]
  {10.1086/519554}, \href
  {https://ui.adsabs.harvard.edu/abs/2007ApJ...665.1413V} {665, 1413}

\bibitem[\protect\citeauthoryear{{Van Eylen} \& {Albrecht}}{{Van Eylen} \&
  {Albrecht}}{2015}]{VanEylen2015}
{Van Eylen} V.,  {Albrecht} S.,  2015, \mn@doi [\apj]
  {10.1088/0004-637X/808/2/126}, \href
  {https://ui.adsabs.harvard.edu/abs/2015ApJ...808..126V} {808, 126}

\bibitem[\protect\citeauthoryear{{Van Eylen}, {Agentoft}, {Lundkvist},
  {Kjeldsen}, {Owen}, {Fulton}, {Petigura}  \& {Snellen}}{{Van Eylen}
  et~al.}{2018}]{VanEylen2018}
{Van Eylen} V.,  {Agentoft} C.,  {Lundkvist} M.~S.,  {Kjeldsen} H.,  {Owen}
  J.~E.,  {Fulton} B.~J.,  {Petigura} E.,   {Snellen} I.,  2018, \mn@doi
  [\mnras] {10.1093/mnras/sty1783}, \href
  {https://ui.adsabs.harvard.edu/abs/2018MNRAS.479.4786V} {479, 4786}

\bibitem[\protect\citeauthoryear{{Vidal-Madjar}, {Lecavelier des Etangs},
  {D{\'e}sert}, {Ballester}, {Ferlet}, {H{\'e}brard}  \&
  {Mayor}}{{Vidal-Madjar} et~al.}{2003}]{VidalMadjar2003}
{Vidal-Madjar} A.,  {Lecavelier des Etangs} A.,  {D{\'e}sert} J.~M.,
  {Ballester} G.~E.,  {Ferlet} R.,  {H{\'e}brard} G.,   {Mayor} M.,  2003,
  \mn@doi [\nat] {10.1038/nature01448}, \href
  {https://ui.adsabs.harvard.edu/abs/2003Natur.422..143V} {422, 143}

\bibitem[\protect\citeauthoryear{{Weiss} \& {Marcy}}{{Weiss} \&
  {Marcy}}{2014}]{Weiss2014}
{Weiss} L.~M.,  {Marcy} G.~W.,  2014, \mn@doi [\apjl]
  {10.1088/2041-8205/783/1/L6}, \href
  {https://ui.adsabs.harvard.edu/abs/2014ApJ...783L...6W} {783, L6}

\bibitem[\protect\citeauthoryear{{Weiss} et~al.,}{{Weiss}
  et~al.}{2018a}]{Weiss2018}
{Weiss} L.~M.,  et~al., 2018a, \mn@doi [\aj] {10.3847/1538-3881/aa9ff6}, \href
  {https://ui.adsabs.harvard.edu/abs/2018AJ....155...48W} {155, 48}

\bibitem[\protect\citeauthoryear{{Weiss} et~al.,}{{Weiss}
  et~al.}{2018b}]{Weiss2019}
{Weiss} L.~M.,  et~al., 2018b, \mn@doi [\aj] {10.3847/1538-3881/aae70a}, \href
  {https://ui.adsabs.harvard.edu/abs/2018AJ....156..254W} {156, 254}

\bibitem[\protect\citeauthoryear{{Wu}}{{Wu}}{2019}]{Wu2019}
{Wu} Y.,  2019, \mn@doi [\apj] {10.3847/1538-4357/ab06f8}, \href
  {https://ui.adsabs.harvard.edu/abs/2019ApJ...874...91W} {874, 91}

\bibitem[\protect\citeauthoryear{{Wu} \& {Lithwick}}{{Wu} \&
  {Lithwick}}{2013}]{Wu2013}
{Wu} Y.,  {Lithwick} Y.,  2013, \mn@doi [\apj] {10.1088/0004-637X/772/1/74},
  \href {https://ui.adsabs.harvard.edu/abs/2013ApJ...772...74W} {772, 74}

\bibitem[\protect\citeauthoryear{{Wyatt}, {Kral}  \& {Sinclair}}{{Wyatt}
  et~al.}{2019}]{Wyatt2019}
{Wyatt} M.~C.,  {Kral} Q.,   {Sinclair} C.~A.,  2019, \mn@doi [\mnras]
  {10.1093/mnras/stz3052}, \href
  {https://ui.adsabs.harvard.edu/abs/2019MNRAS.tmp.2660W} {p.~2660}

\bibitem[\protect\citeauthoryear{{Xie}}{{Xie}}{2014}]{Xie2014}
{Xie} J.-W.,  2014, \mn@doi [\apjs] {10.1088/0067-0049/210/2/25}, \href
  {https://ui.adsabs.harvard.edu/abs/2014ApJS..210...25X} {210, 25}

\bibitem[\protect\citeauthoryear{Zeng et~al.,}{Zeng et~al.}{2019}]{Zeng2019}
Zeng L.,  et~al., 2019, \mn@doi [Proceedings of the National Academy of
  Sciences] {10.1073/pnas.1812905116}, 116, 9723

\bibitem[\protect\citeauthoryear{{Zink}, {Christiansen}  \& {Hansen}}{{Zink}
  et~al.}{2019}]{Zink2019}
{Zink} J.~K.,  {Christiansen} J.~L.,   {Hansen} B. M.~S.,  2019, \mn@doi
  [\mnras] {10.1093/mnras/sty3463}, \href
  {https://ui.adsabs.harvard.edu/abs/2019MNRAS.483.4479Z} {483, 4479}

\makeatother
\end{thebibliography}


\bsp	
\label{lastpage}
\end{document}